\documentclass[aps,prx,twocolumn]{revtex4-2}
\usepackage{graphicx}
\usepackage{amsmath}
\usepackage{subcaption} 
\usepackage{rotating}
\usepackage{color}
\usepackage{hyperref}
\begin{document}

%\preprint{}

\title{ResoSeg: Resonance Tagger using Transformer and Segment Model}

\author{Chunkai Li}
\affiliation{Nankai University, Tianjin, 300071, China}

\author{Ke Li}
\email{like@ihep.ac.cn}
\affiliation{Institute of High Energy Physics, Chinese Academy of Sciences, Beijing, 100049, China}
\affiliation{University of Chinese Academy of Sciences, Beijing, 100049, China.}

\author{Jingde Chen}
\affiliation{Institute of High Energy Physics, Chinese Academy of Sciences, Beijing, 100049, China}
\affiliation{University of Chinese Academy of Sciences, Beijing, 100049, China.}

\author{Junhao Yin}
\email{yinjh@nankai.edu.cn}
\affiliation{Nankai University, Tianjin, 300071, China}

\date{\today}

\begin{abstract}
Deep learning has been widely applied across many areas of experimental high-energy physics, yet existing models address only event-level classification or object tagging and therefore still require reconstruction algorithms tailored to each decay channel. We present the first application of segmentation to resonance tagging at BESIII and introduce ResoSeg, a deep learning model that jointly performs particle-level segmentation and event-level classification, enabling a one-pass analysis of resonance to anything decays while precisely reconstructing the relevant resonance properties. We demonstrate the reconstruction of  $\eta_c$ with $e^+e^-\to\pi^+\pi^-h_c$,\;$h_c\to\gamma\eta_c$,\;$\eta_c\to\text{anything}$. The model is trained on BESIII-$\eta_c$ dataset, which is constructed with per-track true labels obtained via a Truth-Matching Algorithm. Experimental results show that the average combined efficiency of ResoSeg is more than double that of the conventional 16-channel approach across energy points from 4.19 to 4.60\,GeV. The model generalizes to unseen energy points, adapts to other $\eta_c$ production modes through transfer learning, and remains robust against variations in the $\eta_c$ mass, width, and branching fractions, providing a general, resonance-aware model applicable beyond $\eta_c$ and BESIII. The source code is available at \url{https://github.com/oashen/ResoSeg}.
\end{abstract}

%\keywords{}

\maketitle

\section{Introduction}

As a fundamental part of the Standard Model, the quark model classifies ordinary hadrons into mesons and baryons, but also predicts exotic states beyond this picture, such as multi-quark states, glueballs, and hybrids~\cite{Brambilla:2019esw}. 
In experimental high-energy physics, most hadrons are unstable and appear as resonant states---short-lived intermediate particles that manifest as peaks in the invariant mass distributions of their decay products. 
The identification and precise reconstruction of these resonant states are essential for mapping the hadron spectrum, testing QCD predictions, and searching for exotic forms of matter. 

Machine learning, particularly deep learning, has seen broad adoption in experimental high-energy physics. Existing models, however, typically focus on event-level classification or object tagging, i.e., determining whether a given event corresponds to the signal process of interest or contains a certain type of jet.
Powerful classifiers such as Particle Transformer (ParT)~\cite{ParT}, SPANet~\cite{SPANet} and GN3~\cite{GN3} have achieved state-of-the-art performance in jet tagging,
and have also been widely applied to event classification at lepton colliders~\cite{ParT_work1,ParT_work2,ParT_work3}.
The study of intermediate resonances is especially important at lepton colliders. Compared with hadron colliders, lepton colliders such as BESIII, Belle II, and CEPC operate with substantially lower background levels; therefore, the key challenge is not background suppression but rather increasing the available signal statistics.
Despite their success, these models do not substantially improve signal statistics significantly in this context: they do not provide per-track information about the origin of particles relative to intermediate resonances.
Consequently, analyses that adopt such models still require dedicated reconstruction algorithms tailored to each specific decay channel to reconstruct the four-momentum of the target resonance, leaving the overall analysis paradigm unchanged.

Segmentation, which assigns a semantic label to each element of an input, offers a natural solution to this limitation. 
When adapted to experimental high-energy physics, track-level segmentation enables the direct identification of which tracks originate from a given intermediate resonance, 
allowing the simultaneous reconstruction of resonance to anything decay chains in a single inference pass, thereby substantially increasing the available data statistics compared with conventional channel-by-channel approaches. 
Crucially, it provides access to the full kinematic properties of the resonance without relying on decay-channel-specific reconstruction algorithms, transforming the way analyses are performed.

The study of hadron spectroscopy and exotic hadron candidates is a central program at multiple experiments, including BESIII, Belle II, and LHCb. A common requirement across these analyses is the identification and reconstruction of intermediate resonances---for example, $\eta_c$ and $J/\psi$ at BESIII, $B$ and $D$ mesons at LHCb, and $B$ mesons at Belle II---whose decay products must be disentangled from the remaining particles in the event. In all cases, the precision of the resonance reconstruction directly determines the sensitivity of the physics measurement.

In this article, we propose ResonanceSegmenter (ResoSeg), a unified model that combines event-level classification with particle-level segmentation. ResoSeg simultaneously determines whether an event contains a signal process and assigns a semantic category to each reconstructed track, thereby identifying the target resonance and labeling its decay products in a single model. Specifically, ResoSeg first embeds the track features into a high-dimensional latent space via an embedding layer, and then segments the tracks within this latent space, thereby determining whether each track originates from the target resonance and reconstructing the resonance.

We focus on the $\eta_c$ meson as our primary demonstration case, but the ResoSeg model is inherently general and can be readily extended to other resonances and experiments. As the lightest ground-state charmonium~\cite{PDG}, $\eta_c$ serves as a fundamental spectroscopic benchmark, yet its reconstruction is notoriously challenging: it possesses hundreds of possible decay channels~\cite{PDG}, most with tiny branching fractions, and no single dominant mode, making channel-by-channel reconstruction impractical. Traditional reconstruction methods typically concentrate on the 16 decay channels with relatively large branching fractions and cleaner backgrounds, and have achieved considerable success~\cite{traditional,traditional2,traditional3}. Building on this success, we aim to further improve the reconstruction performance with ResoSeg to obtain even better results.

\section{BESIII detector}
The BESIII detector~\cite{Ablikim:2009aa} records symmetric $e^+e^-$ collisions 
provided by the BEPCII storage ring~\cite{Yu:IPAC2016-TUYA01}
in the center-of-mass energy range from 1.84 to 4.95~GeV,
with a peak luminosity of $1.1 \times 10^{33}\;\text{cm}^{-2}\text{s}^{-1}$ 
achieved at $\sqrt{s} = 3.773\;\text{GeV}$. 
Large data samples have been collected in this energy region~\cite{Ablikim:2019hff}. The cylindrical core of the BESIII detector covers 93\% of the full solid angle and consists of a helium-based
multilayer drift chamber~(MDC), a time-of-flight
system~(TOF), and a CsI(Tl) electromagnetic calorimeter~(EMC),
which are all enclosed in a superconducting solenoidal magnet
providing a 1.0~T magnetic field.
The solenoid is surrounded by an octagonal flux-return yoke made of steel, interleaved with resistive-plate-counter muon-identification modules.
%The acceptance of charged particles and photons is 93\% over $4\pi$ solid angle. 
The charged-particle momentum resolution at $1~{\rm GeV}/c$ is
$0.5\%$, and the 
${\rm d}E/{\rm d}x$
resolution is $6\%$ for electrons
from Bhabha scattering. The EMC measures photon energies with a
resolution of $2.5\%$ ($5\%$) at $1$~GeV in the barrel (end cap)
region. The time resolution in the plastic scintillator TOF barrel region is 68~ps, while
that in the end cap region was 110~ps. 
The end cap TOF
system was upgraded in 2015 using multigap resistive plate chamber
technology, providing a time resolution of
60~ps,
which benefits 33\% of the data used in this analysis~\cite{etof1,etof2,etof3}.

Since it began operation, BESIII has played a leading role in the discovery and investigation of exotic states, also known as the {\it XYZ} states~\cite{Yuan_review}. More than 30 new particles have been obseved by BESIII experiment~\cite{PDG}. Many of the new particles are observed from the reconstruction of final states involving $\eta_c$, for example, $Z_c(3900) \to \pi^+ h_c\;[h_c \to \gamma \eta_c],\;\rho^+ \eta_c$, etc~\cite{4230_pipihc,3900_pihc,3res_pipihc,pi0pi0hc,3pietac}.

%Many of these searches have been carried out by reconstructing the $\eta_c$ meson, for example, the observation of several resonances through $Y\to\pi^+Z_c,\;Z_c\to\pi^-\gamma \eta_c$ decays~\cite{4230_pipihc,3900_pihc,3res_pipihc,pi0pi0hc} and through $e^+e^-\to\pi^+\pi^-\pi^0\eta_c$~\cite{3pietac}. These studies highlight the importance of $\eta_c$ reconstruction as a powerful tool for exotic hadron spectroscopy, and motivate the development of more efficient and general-purpose reconstruction methods such as the one proposed in this work. 

%\input{Realated_work}

\section{BESIII-$\eta_c$ Dataset}
\label{Sec:dataset}

In this section, we provide an overview of the BESIII-$\eta_c$ dataset. 
In this work, the signal channel is $e^+e^-\to\pi^+\pi^-h_c$,\;$h_c\to\gamma \eta_c$,\;$\eta_c\to \text{anything}$; the background channel is $e^+e^-\to\pi^+\pi^-h_c$,\;$h_c\to \text{anything}$ except $\gamma \eta_c$.

The dataset comprises the inclusive decay modes of the $\eta_c$ resonance. Each event includes an event-level true label that indicates whether the event is a signal event, and particle-level true labels that classify all reconstructed tracks from the BESIII detector into three categories:
\begin{enumerate}
	\item fake tracks or background-event tracks;
	\item real tracks originating from the intermediate $\eta_c$ resonance;
	\item real tracks not originating from the intermediate $\eta_c$ resonance (e.g., the $\pi^+\pi^-$ from $e^+e^-$ annihilation, and the $\gamma$ from $h_c$ decay).
\end{enumerate}

\textbf{Simulation setup.} The Monte Carlo (MC) dataset is simulated with standard Monte Carlo event generators used by the BESIII experiment. All particle decays are modeled with EVTGEN~\cite{EVTGEN} using branching fractions
either taken from the
Particle Data Group (PDG)~\cite{PDG}, when available,
or otherwise estimated with LUNDCHARM~\cite{LUNDCHARM}.\\

\begin{figure}
	%\begin{subfigure}[b]{0.45\textwidth}
		%\centering
		\includegraphics[width=0.9\columnwidth]{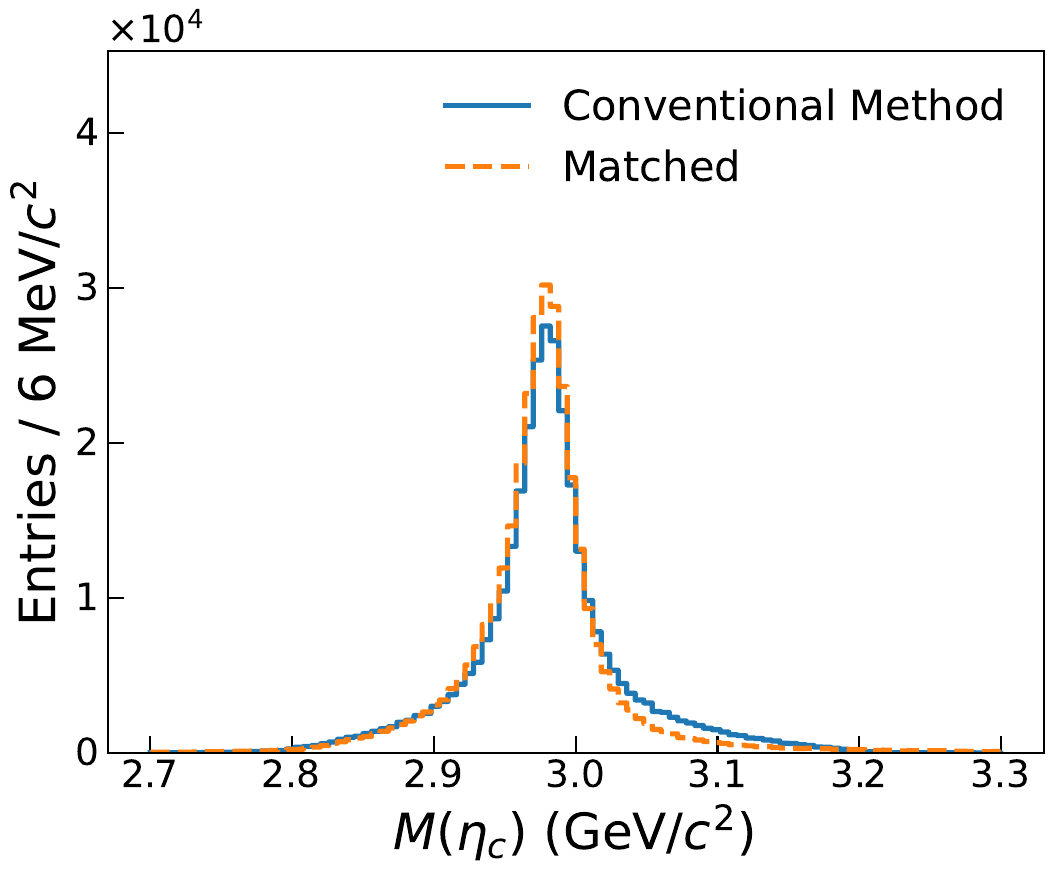}
	%\end{subfigure}\\
	%\begin{subfigure}[b]{0.45\textwidth}
	%	%\centering
	%	\includegraphics[width=\textwidth]{pics/R.pdf}
	%	\caption{Mass resolution}   % 自动生成 (b)
	%	\label{fig:sub2}
	%\end{subfigure}
	\caption{\raggedright \label{fig:etac_mass} The matched $\eta_c$ invariant mass spectrum compared with the distribution obtained by the conventional method.}
\end{figure}

\textbf{True label.} %Since the BESIII simulation algorithm does not directly provide particle-level true labels, 
The true labels needed for supervised learning are obtained via a Truth-Matching Algorithm. In this algorithm, we define an average invariant mass resolution $\rm Res$ to evaluate the quality of the truth matching,
\begin{equation}
	{\rm Res}=\sum_{\rm resonances} \frac{|M_{\rm rec}-M_{\rm truth}|}{M_{\rm truth}} .
\end{equation}
The best matching result is the one with the minimum $\rm Res$ value. The matched $\eta_c$ invariant mass spectrum obtained with this method is shown in Fig.~\ref{fig:etac_mass}, together with the distribution obtained by the conventional method. The good agreement indicates that the Truth-Matching Algorithm successfully obtains the true labels. The efficiency of the Truth-Matching Algorithm varies with the decay channel and the center-of-mass energy. For the decay channel $e^+e^-\to \pi^+\pi^- h_c$,\;$h_c\to \gamma \eta_c$,\;$\eta_c \to \text{anything}$, the average efficiency at the center-of-mass energies of 4.22\,GeV, 4.26\,GeV, and 4.39\,GeV is 5.68\%.

\textbf{Input features.} Information about the charged and neutral tracks in each event is taken as the input. The number of tracks per event is not fixed, but generally does not exceed 50, with the majority distributed in the range of 7 to 15. Charged tracks with sufficiently large momentum typically leave responses in the MDC, TOF, and EMC. Their responses in the MDC allow us to reconstruct their trajectories as helices, while those in the TOF and EMC, together with the MDC response, help with particle identification (PID). Neutral tracks (usually photons) leave responses only in the EMC and can be reconstructed as electromagnetic showers. For each track, the following three types of information are chosen: MDC track information, EMC shower information, and PID information.
\begin{itemize}
	\item \textbf{MDC track information.} This item contains the charge $q$, the distances $v_r$, $v_z$ from the reconstructed track to the nearest point of the interaction point, and the five helix parameters ($d_0,\;\phi_0 ,\;\kappa ,\;z_0 ,\; \tan \lambda$) of each track.
	\item \textbf{EMC shower information.} This item contains the polar angle $\theta$ and azimuthal angle $\phi$ describing the shower position, the second moment $A_{20}$ and lateral moment $A_{42}$ characterizing the shower shape, and the total deposited energy $E$ in the EMC, along with their uncertainties $d\theta$, $d\phi$, and $dE$.
	\item \textbf{PID information.} This item contains the PID classification results obtained from the BESIII PID algorithm. The class of a charged track is determined as the hypothesis with the highest probability output by the PID algorithm. The label for neutral tracks is 0, while for charged tracks, the labels for the $e$, $\mu$, $\pi$, $K$, $p$ hypotheses are 1, 2, 3, 4, and 5, respectively.
\end{itemize}

\textbf{Training, validation and test sets.} The dataset includes three center-of-mass energy points: 4.22\,GeV, 4.26\,GeV, and 4.39\,GeV. For each energy point, 0.8 million events are generated for both the signal channel and the background channel, resulting in a total of 4.8 million events. They are split into training, validation, and test sets in a ratio of 8:1:1. 

Furthermore, to validate the generalization ability of the model and its performance in realistic physics analysis scenarios, we also use the official BESIII inclusive MC dataset at 4.26 GeV as a test set. This dataset contains events of all known hadronic decay processes of $e^+e^-$. This inclusive MC dataset is required to satisfy the following preselections:
\begin{enumerate}
	\item At least one set of $\gamma$, $\pi^+$, $\pi^-$ candidates satisfies $3.45\ \rm{GeV}/c^2 < RM(\pi^+\pi^-) < 3.65\ \rm{GeV}/c^2$ and $2.80\ \rm{GeV}/c^2 < RM(\gamma\pi^+\pi^-) < 3.20\ \rm{GeV}/c^2$;
	\item The event must be correctly and fully reconstructed, i.e., passing a kinematic fit to itself.
\end{enumerate}
After the preselections, the remaining number of events is 222,125.

\textbf{Evaluation metrics.} The labeling of intermediate resonances can be essentially divided into two tasks: event-level classification and particle-level segmentation. For event-level classification, we use accuracy Acc and recall as evaluation metrics. For particle-level segmentation, we also need to evaluate the model's performance at the event level. To this end, we adopt the IoU score, as defined in Eq.~\ref{IoU}, which is commonly used in image segmentation. The IoU score is an event-level metric with values in the range $[0,1]$. The closer the value is to 1, the closer the model's output matches the true labels for that event. We quantify segmentation performance using the average IoU score per category and the event-level perfect match rate (PMR), which is defined as the fraction of events with a perfect IoU score of 1 for all categories.
\begin{equation}
	\label{IoU}
	\rm{IoU} = \frac{\rm{Model\; Output}\cap \rm{True\;Label}}{\rm{Model\; Output}\cup \rm{True\;Label}},
\end{equation}
Combining both aspects, we use the model signal reconstruction efficiency $\epsilon_{\rm rel,\;sig} = \frac{N_{\rm final}}{N_{\rm TM}}$ (i.e., the relative efficiency of the model), where $N_{\mathrm{TM}}$ denotes the number of signal events passing the Truth-Matching Algorithm, and $N_{\mathrm{final}}$ is the number of signal events correctly reconstructed after model inference.

For the inclusive MC test set, we use the background rejection rate $\rm{Rej}=1/\epsilon_{B}$ to evaluate the model performance, where $\epsilon_{B}$ is the background mistag efficiency.

\section{Model Architecture}
\begin{figure*}
	\begin{subfigure}[b]{0.9\textwidth}
		%\centering
		\includegraphics[width=\textwidth]{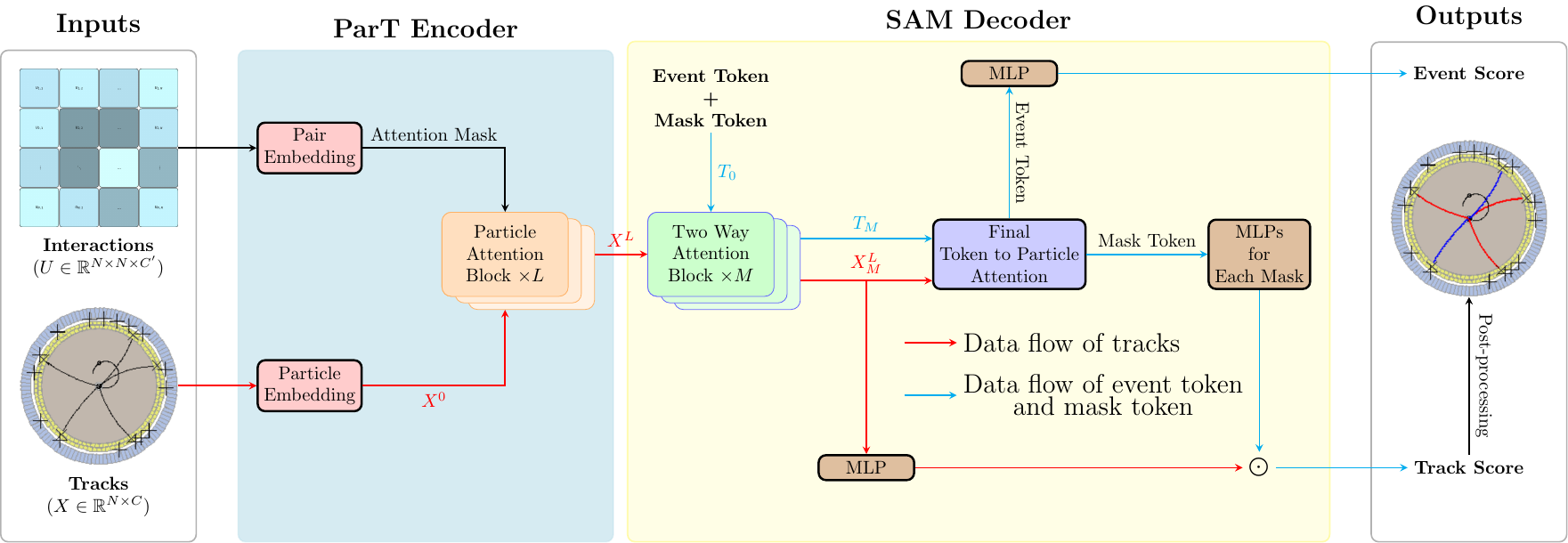}
		\caption{ResoSeg}
	\end{subfigure}
	\begin{subfigure}[b]{0.9\textwidth}
		%\centering
		\includegraphics[width=\textwidth]{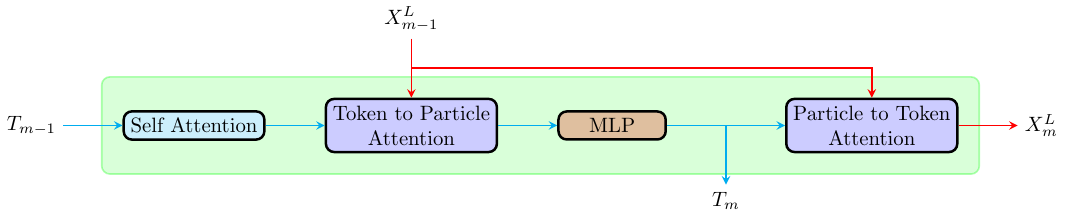}
		\caption{Two Way Attention Block}
	\end{subfigure}
	\caption{\raggedright \label{fig:full architecture} The architecture of (a) ResoSeg (b) Two Way Attention Block.}
\end{figure*}
An overview of the ResoSeg architecture is presented in Fig.~\ref{fig:full architecture}. The encoder of the model consists of Particle Embedding, Pair Embedding, and Particle Attention Blocks, drawing inspiration from ParT~\cite{ParT}. The decoder is inspired by SAM~\cite{SAM}, an image segmentation model. Similar to ParT, since the input tracks are permutation invariant under token exchange, ResoSeg does not introduce any positional encoding. Additionally, it introduces interaction input features as an inductive bias to participate in the self-attention computation among the input particles. For an event with $N$ tracks (including both charged and neutral tracks), the Tracks input includes a list of $C$ features for every track and forms a shape $(N,C)$, while the Interactions input is a matrix of $C'$ features for every pair of particles with shape $(N,N,C')$. For the interaction features used to describe track-pair information, we adopt several features that differ from those used in ParT; these features are optimized by Bayesian optimization~\cite{bayes}, which is discussed in Sec.~\ref{Sec:bayes}. Specifically, for a pair of tracks $a$, $b$ with 4-vectors $p_a=(E_a,\vec{P_a})$ and $p_b=(E_b,\vec{P_b})$, charges $q_a$ and $q_b$,
%and deposit energy in EMC $E_{a,EMC}$, $E_{b,EMC}$,
we calculate the following four features:
\begin{equation}
	\begin{split}
	&M^2=(E_a+E_b)^2-|\vec{P_a}+\vec{P_b}|^2,\\
	&RM^2=(E_{\rm cms}-E_a-E_b)^2-|\vec{P_{\rm cms}}-\vec{P_a}-\vec{P_b}|^2,\\
	&Q = q_a*q_b,\\
	&k_t = min(P_{T,a},P_{T,b})\sqrt{(y_a-y_b)^2+(\phi_a-\phi_b)^2},
	\end{split}
\end{equation}
where $P_{T,i}=(P_{x,i}^2+P_{y,i}^2)^{1/2}$ is the transverse momentum, for $i = a,\;b$ and $E_{\rm cms}$ is the center-of-mass energy. In particular, $RM^2$ explicitly includes the event-level center-of-mass energy and momentum, thereby providing global kinematic context that is not captured by the individual track features alone. Since $M^2$, $RM^2$ and $k_t$ typically have a long-tail distribution, we use their logarithmic forms.

The tracks and interactions inputs are each followed by an MLP that projects them to $d$- and $d'$-dimensional embeddings, which are then fed into the Particle Attention Blocks to produce the encoded latent representations $X^L$ that are passed to the SAM Decoder. The outputs of the SAM Decoder are the score for the entire event and the score for each track. The event score indicates how likely the event is to be a signal event, while the track score indicates the category of the track.

\textbf{Particle Attention Block.} The Particle Attention Block is essentially the same as the one used in ParT, except that the activation function is replaced from GELU to SwiGLU. Each Particle Attention Block consists of a P-MHA, a two-layer MLP, and residual connections~\cite{ParT}.

\textbf{SAM Decoder.} The SAM Decoder is the core component of ResoSeg. The SAM Decoder we use follows the original SAM design~\cite{SAM} closely, differing only in its inputs and outputs. Regarding the inputs, $X^L$ is the sole input, without any prompt tokens. Regarding the outputs, the original SAM performs a binary classification task, while ours performs a multi-class classification task.

In image segmentation, a mask is a pixel-wise label map that assigns a semantic class to every pixel. In ResoSeg, the segmentation target is a set of tracks instead of pixels, so the mask corresponds to assigning a semantic class to each track. The Mask Token is a learnable embedding that encodes this track-level mask information and interacts with particle features to generate the final track scores.

For each event, we randomly initialize the Event Token and Mask Token, both of which have shape $(1,d)$. These tokens pass through $M$ layers of Two Way Attention Block together with $X^L$. As in SAM, each Two Way Attention Block includes a self-attention on the Event Token and Mask Token (except for the first layer) and two cross-attentions between these two tokens and the Particle Embedding. After all Two Way Attention Block layers, the tokens undergo one more token-to-particle attention as a final update. Subsequently, the Event Token is passed through an MLP to output the Event Score; the Mask Token and the updated $X^L_M$ pass through their own MLPs, respectively, reducing their dimensions to $d'' \times N_{\text{class}}$ and $d''$. Here, $N_{\text{class}}$ is the number of categories for the particle-level segmentation. Then, a reshape operation transforms the Mask Token of shape $(1, d'' \times N_{\text{class}})$ into shape $(N_{\text{class}}, d'')$ to enable multi-class classification. Finally, a point-wise product is performed between the Mask Token and $X^L_M$ to produce the track score output.
 
 \textbf{Losses.} Both the event-level classification and particle-level segmentation tasks use cross-entropy as the loss. The total loss is a linear combination of the two losses.
\begin{equation}
	\mathcal{L} = f_1 \times \mathcal{L}_{event} + f_2 \times \mathcal{L}_{particle}
\end{equation}
 
\textbf{Implementation.} We implement the ResoSeg model in PyTorch~\cite{pytorch}. The baseline model has a total of $L=4$ Particle Attention Blocks and $M=2$ Two Way Attention Blocks. It uses a particle embedding of dimension $d = 64$, encoded from the input particle features using a three-layer MLP with $(64, 256, 64)$ nodes per layer, with GELU nonlinearity and layer normalization (LN) in between. The interaction input features are encoded using a four-layer pointwise 1D convolution with $(64, 64, 64, 4)$ channels, with GELU nonlinearity and batch normalization in between, to yield a $d'=8$-dimensional interaction matrix. All MHA blocks, including the P-MHA in the Particle Attention Block and the self-attention and cross-attention in the SAM Decoder, have 8 heads and an expansion factor of 4 for the MLP. For the particle-level segmentation task, the number of classification categories is $N_{\text{class}}=3$, corresponding to the three true-label categories provided in the dataset. The MLP before the point-wise product reduces the dimensions of both the Mask Token and the Particle Embedding to $d'' = d/4 = 16$. We use a dropout rate of 0.1 for all attention blocks. We also investigated using uncertainty weighting~\cite{UW} to dynamically adjust the loss coefficients $f_1$ and $f_2$, but found no significant improvement over a fixed equal weighting; therefore, we simply set $f_1 = f_2 = 0.5$. We further tested gated attention (GA)~\cite{gated_attn} and conditional layer normalization (CLN)~\cite{CLN}, but neither led to a noticeable performance gain; the detailed results are presented in Sec.~\ref{Sec:gate}. The hyperparameters were tuned within a reasonable range, providing a solid baseline. The total number of learnable parameters in the model is 557 thousand.

\section{Experiments}
We conduct experiments on the BESIII-$\eta_c$ dataset and present the results in Sec.~\ref{Sec:base}. The dataset is designed for the specific process described in Sec.~\ref{Sec:dataset}. ResoSeg trained on this dataset serves as a pre-trained model. We then adopt a transfer learning approach by changing the target decay channel to $\psi(2S)\to\gamma\eta_c(2S)$,\;$\eta_c(2S)\to\gamma h_c$,\;$h_c\to\gamma\eta_c$ to evaluate its generalization capability to other $\eta_c$ decay channels; the results are shown in Sec.~\ref{Sec:fine-tuning}. We perform robustness tests against variations in the $\eta_c$ mass, width, and branching fractions, with the results presented in Sec.~\ref{Sec:robustness}.
% Maybe in discussion?
%Based on the aforementioned model as a baseline{\color{red} (redundant baseline?)}, we introduce some architectural modifications and robustness evaluation. The details and results are presented in Section \ref{Sec:test}. 

\subsection{Result on the BESIII-$\eta_c$ dataset}
\label{Sec:base}
\subsubsection{Setup}
We use the full BESIII-$\eta_c$ dataset as the training data and optimize the parameters with the Adam~\cite{Adam} optimizer. The batch size is 1024, and the model is trained for 100 epochs. The learning rate follows a linear warmup and cosine annealing schedule: the first 20\% of epochs are used for warmup, and the last 10\% for annealing, with a stable learning rate of 0.0001. At the end of each epoch, the model is evaluated on the validation set and a checkpoint is saved. The checkpoint with the lowest validation loss is then used to evaluate the model's performance on the test set.

\subsubsection{Post-processing}
After obtaining the model predictions,
% to achieve the best particle-level segmentation performance, we optimize the track score cut using the mean IoU score, with the optimization objective of maximizing the mean IoU score. The optimization results are shown in Fig \ref{fig:opt_iou}. After obtaining the score cut,
all tracks can be classified into three categories. To better reconstruct the $\eta_c$ resonance and the entire event, we perform the following operations on these tracks: for tracks identified as real tracks from $\eta_c$, we sum their four-momenta to reconstruct the $\eta_c$ resonance; for those identified as real tracks not from $\eta_c$, we require at least one neutral track and two charged tracks, and find the best $\gamma\pi^+\pi^-$ combination by minimizing $|M(\gamma\pi^+\pi^-\eta_c) - M_{\text{cms}}|$.
%\begin{figure}
%	\includegraphics[width=0.45\textwidth]{pics/default_sig_IoU.pdf}
%	\includegraphics[width=0.45\textwidth]{pics/default_gam_IoU.pdf}
%	\caption{\label{fig:opt_iou} The IoU score optimization. The score cut is set to ``real tracks from $\eta_c$'' score $ > 0.5$ and ``real tracks not from $\eta_c$'' score $ > 0.4$.}
%\end{figure}
\subsubsection{Result}
\begin{figure}
	\begin{subfigure}[b]{0.45\textwidth}
		%\centering
		\includegraphics[width=0.9\textwidth]{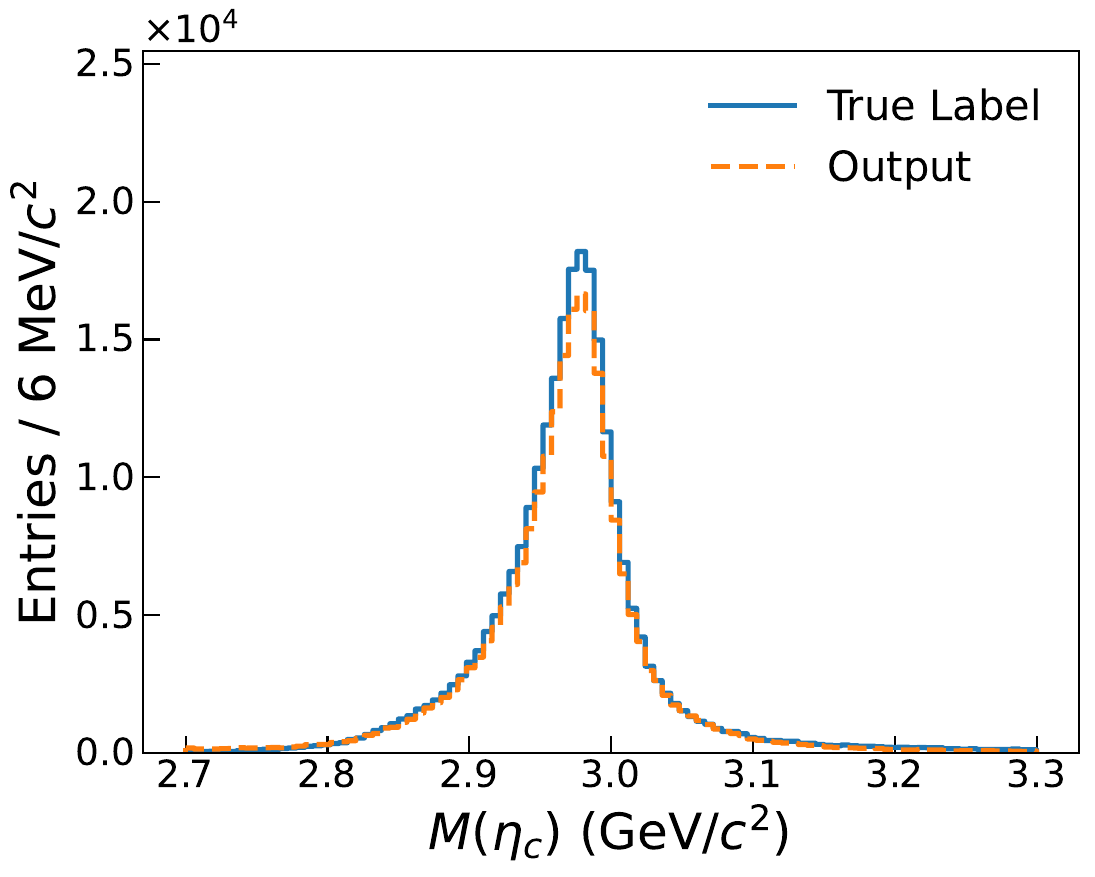}
		\caption{$\eta_c$}
		\label{fig:etac_m_sub1}
	\end{subfigure}\\
	\begin{subfigure}[b]{0.45\textwidth}
		%\centering
		\includegraphics[width=0.9\textwidth]{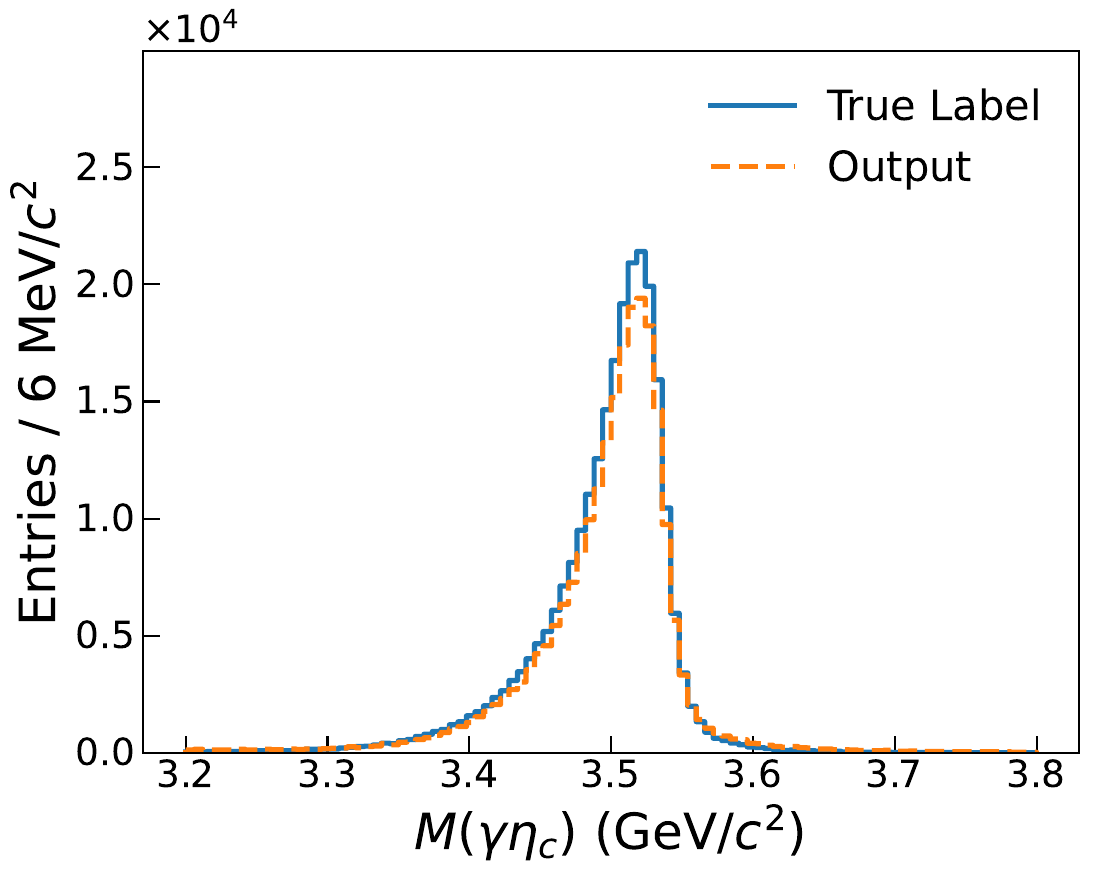}
		\caption{$\gamma\eta_c$}
		\label{fig:etac_m_sub2}
	\end{subfigure}
	\caption{\raggedright \label{fig:etac_m}The reconstructed mass spectra of $\eta_c$ and $\gamma\eta_c$ on the test set, compared with those derived from the true labels.}
\end{figure}

The reconstructed mass spectra of $\eta_c$ and $\gamma\eta_c$ on the test set, together with those derived from the true labels, are shown in Fig.~\ref{fig:etac_m}. The ResoSeg outputs reproduce the genuine signal shapes well.
Performance is evaluated using the metrics described in Sec.~\ref{Sec:dataset}.
For benchmarking purposes, a ParT model with identical hyperparameters is trained on the same dataset, and the results of both models are summarized in Table~\ref{tab:base_metrics}. Here, the reported $\epsilon_{\rm rel,\;sig}$ is the signal efficiency obtained after imposing the $\eta_c$ and $h_c$ mass-window selections together with the event score requirement; the specific selections are $2.85\;\mathrm{GeV}/c^2<M(\eta_c)<3.05\;\mathrm{GeV}/c^2$, $3.40\;\mathrm{GeV}/c^2<M(\gamma\eta_c)<3.58\;\mathrm{GeV}/c^2$, and event score $>0.9$. In contrast, ${\rm Rej}_{90}$ is evaluated using only the event score selection, at the event score threshold that yields a relative signal efficiency of 90\%.
ResoSeg achieves strong performance across all metrics. In event-level classification, the high Acc and recall indicate that the model can effectively distinguish signal from background events. In particle-level segmentation, the high PMR shows that the model reconstructs the majority of signal events exactly,
while the high average IoU indicates that the vast majority of both ``real tracks from $\eta_c$'' and ``real tracks not from $\eta_c$'' are correctly classified.
The favorable $\epsilon_{\rm rel,\;sig}$ and ${\rm Rej}_{90}$ further ensure that ResoSeg retains as much signal as possible while suppressing background effectively.

\begin{figure}
	%\centering
	\includegraphics[width=0.9\columnwidth]{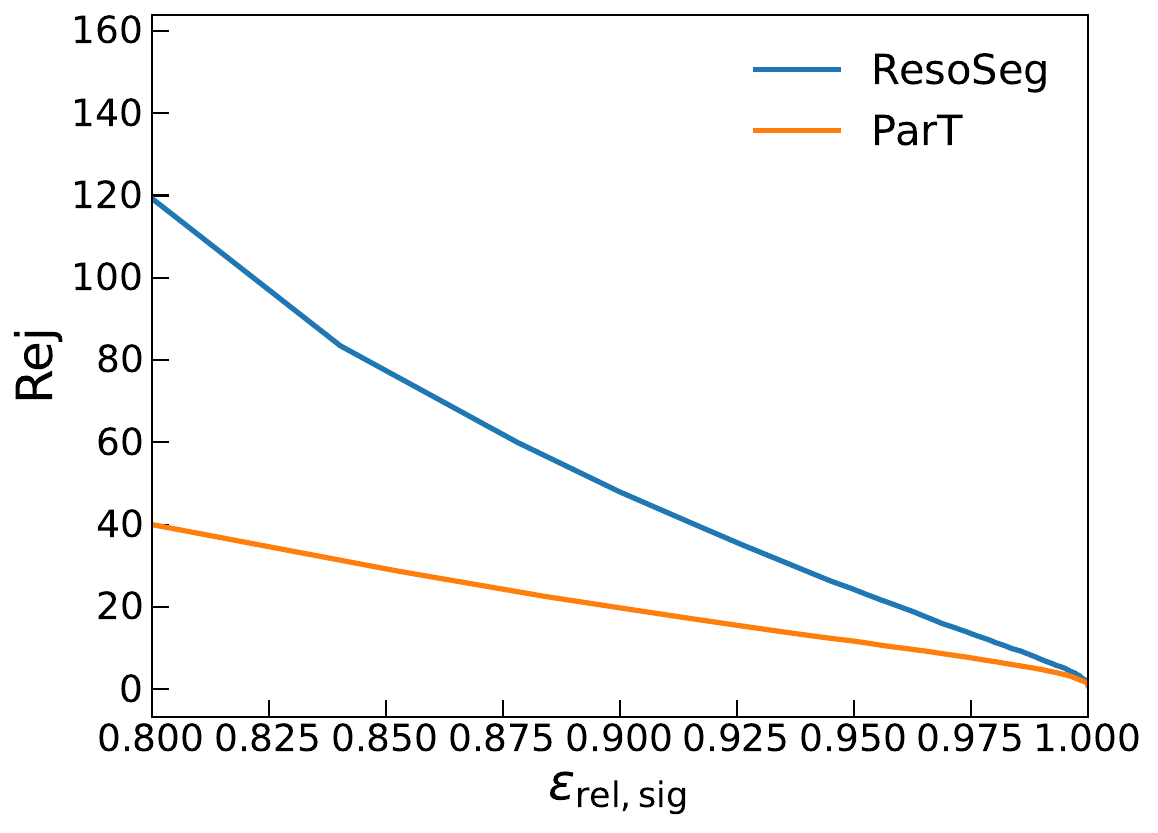}
	\caption{\raggedright \label{fig:eff_rej}The two-dimensional comparison of $\epsilon_{\rm rel,\;sig}$ and Rej for ResoSeg and ParT, where $\epsilon_{\rm rel,\;sig}$ is evaluated using only the event score selection.}
\end{figure}

Within the BESIII-$\eta_c$ dataset, ResoSeg and ParT exhibit comparable performance at the event level. On the more complex inclusive MC test set, ResoSeg shows a clear advantage in the two-dimensional $\epsilon_{\rm rel,\;sig}$ versus Rej comparison, where $\epsilon_{\rm rel,\;sig}$ is evaluated using only the event score selection, as shown in Fig.~\ref{fig:eff_rej}. The results demonstrate that ResoSeg can achieve a substantially higher Rej while maintaining competitive signal efficiency. More importantly, ResoSeg can recover the full four-momentum information of both the $\eta_c$ resonance and $h_c$ resonance, providing further potential for background suppression. This suggests that ResoSeg offers a greater advantage in the analysis of resonance to anything decays.

%Notably, the fact that $\epsilon_{\rm rel}$ is slightly higher than the perfect match rate confirms the effectiveness of the post-processing procedure in recovering otherwise misclassified events.
\begin{table*}
	\caption{\raggedright \label{tab:base_metrics}The evaluation metrics of ParT and ResoSeg on the BESIII-$\eta_c$ dataset and the inclusive MC test set. For the average IoU, results are presented as ``real tracks from $\eta_c$'' / ``real tracks not from $\eta_c$''. The $\epsilon_{\rm rel,\;sig}$ is the signal efficiency obtained after imposing the $\eta_c$ and $h_c$ mass-window selections together with the event score requirement, while the ${\rm Rej}_{90}$ is evaluated using only the event score selection, at the event score threshold that yields a relative signal efficiency of 90\%.}
	\begin{ruledtabular}
		\begin{tabular}{lcccccc}
			% Lines of table here ending with \\
			&		Acc(\%)&	Recall(\%)&		Average IoU($\times 10^{-2}$)&	PMR(\%)&		$\epsilon_{\rm rel,\;sig}$(\%)&	${\rm Rej}_{90}$\\\hline
			ParT&	95.93&		98.93&			\textbackslash&					\textbackslash&	\textbackslash&				19.69\\
			ResoSeg&97.31&		98.84&			96.12/95.20&					87.91&			79.79&						47.82
		\end{tabular}
	\end{ruledtabular}
\end{table*}

\begin{figure}
	%\centering
	\includegraphics[width=0.9\columnwidth]{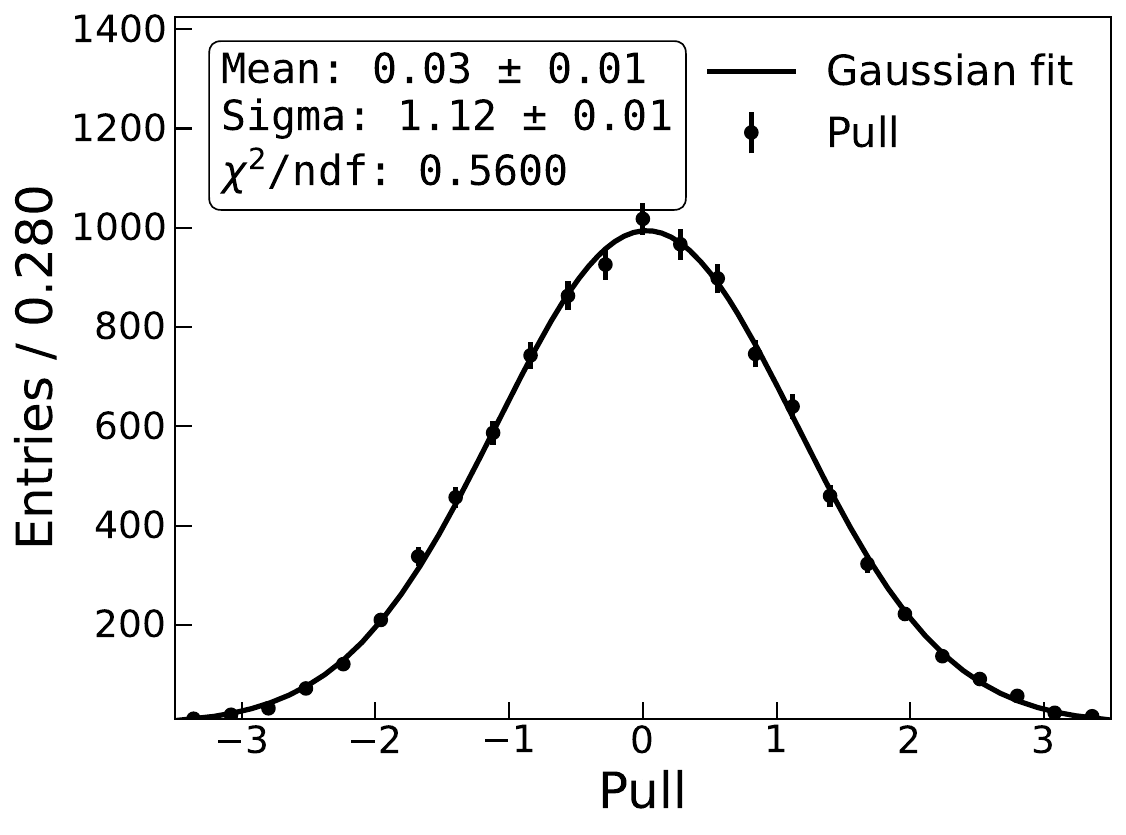}
	\caption{\raggedright \label{fig:io_check}The ${\rm pull}=(N_{\text{input}}-N_{\text{output}})/\sigma_{\text{output}}$ distribution of the input/output check, fitted with a Gaussian function.}
\end{figure}

\begin{figure}
	%\centering
	\includegraphics[width=0.9\columnwidth]{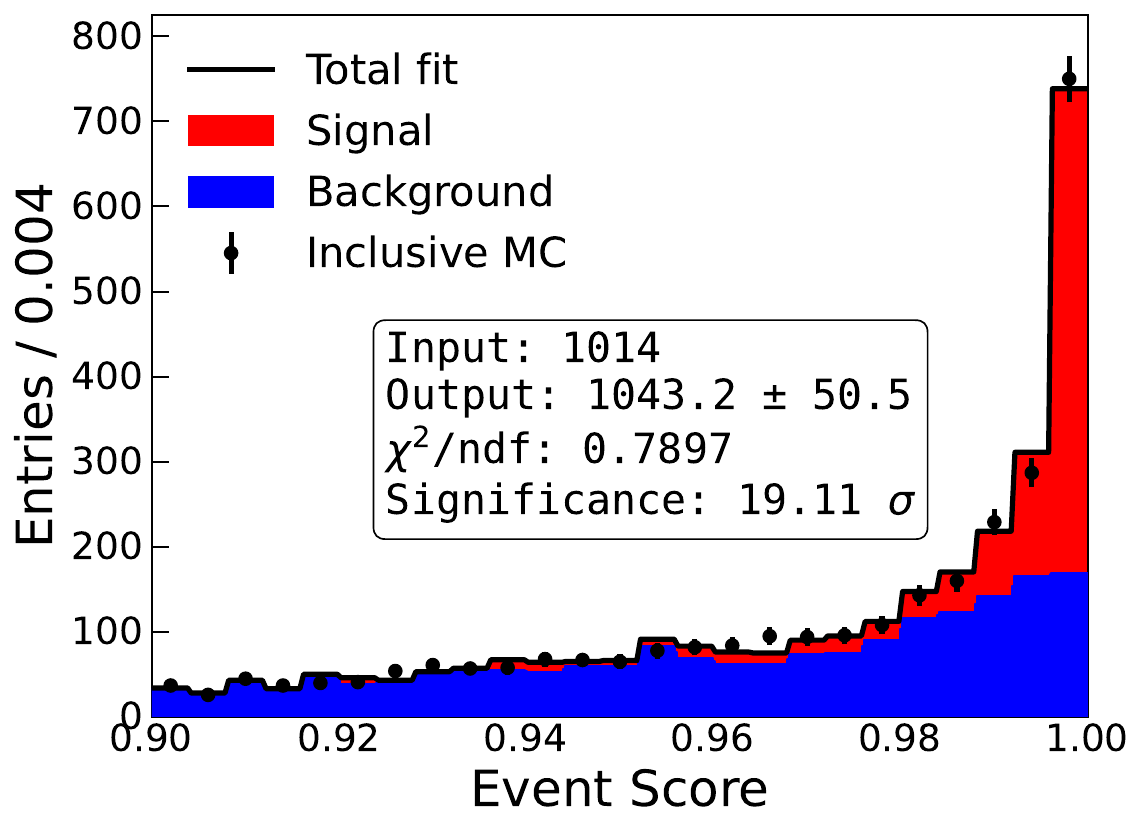}
	\caption{\raggedright \label{fig:fit} One representative fit from the input/output check.}
\end{figure}

\textbf{Signal extraction on inclusive MC test set.} We extract the signal from the event score distribution. The inclusive MC sample is randomly split into two parts: one part is used to extract the probability density functions (PDFs), and the other part is used to fit and extract the event yield. This process is repeated 10,000 times to perform an input/output check. The resulting ${\rm pull}=(N_{\text{input}}-N_{\text{output}})/\sigma_{\text{output}}$ distribution is shown in Fig.~\ref{fig:io_check}, and the fit result from one of these iterations is shown in Fig.~\ref{fig:fit}. The pull distribution is fitted with a Gaussian function, yielding a mean of $0.03 \pm 0.01$ and a $\sigma$ of $1.12 \pm 0.01$, indicating that our method extracts the signal from the inclusive MC sample stably and accurately. Moreover, the fits yield extracted signals of high statistical significance.

\begin{figure}
	\begin{subfigure}[b]{0.45\textwidth}
		%\centering
		\includegraphics[width=\textwidth]{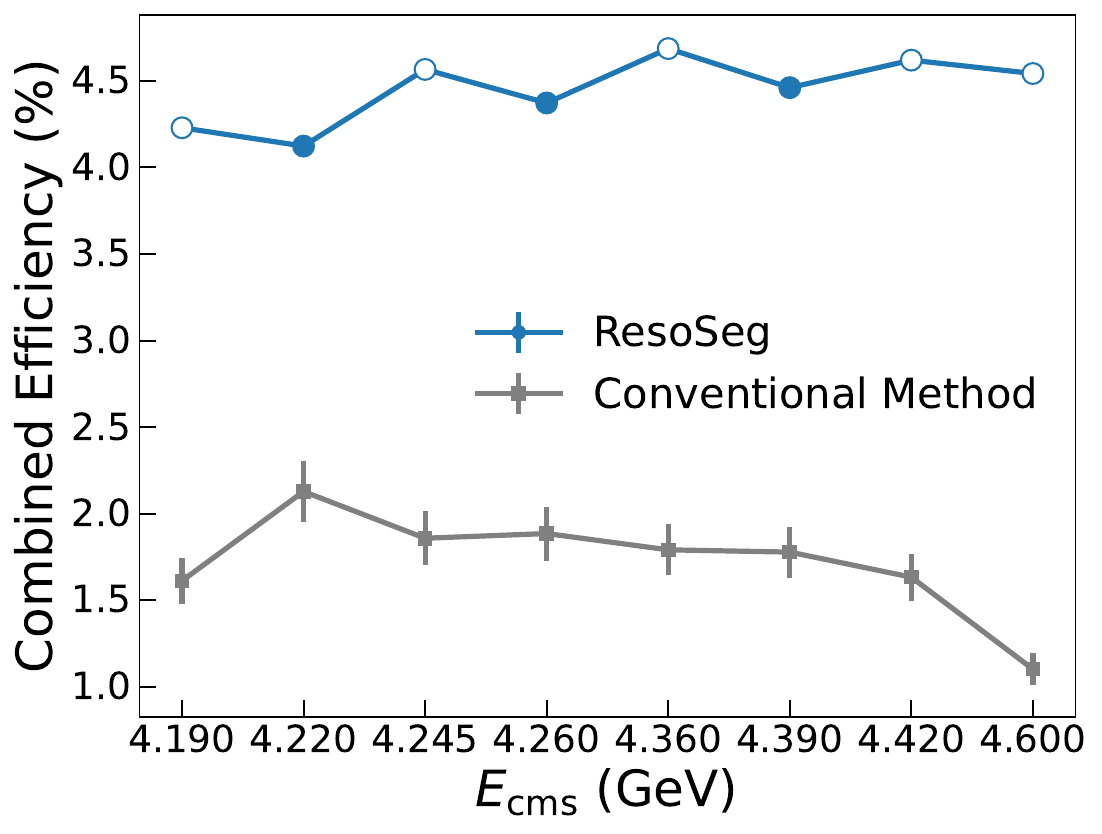}
		\caption{The combined efficiency}
		\label{fig:ecms_sub1}
	\end{subfigure}\\
	\begin{subfigure}[b]{0.45\textwidth}
		%\centering
		\includegraphics[width=\textwidth]{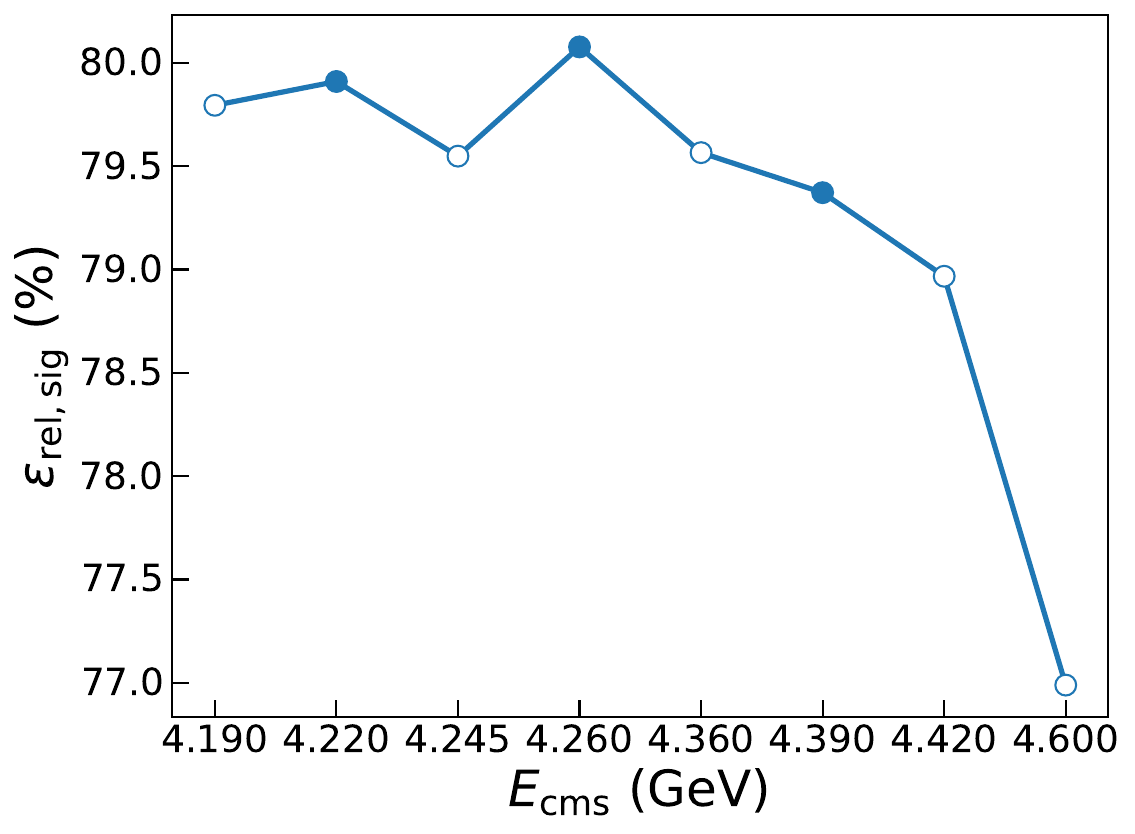}
		\caption{The ResoSeg efficiency $\epsilon_{\rm rel,\;sig}$}
		\label{fig:ecms_sub2}
	\end{subfigure}
	\caption{\raggedright \label{fig:ecms_default} The combined efficiency and the ResoSeg efficiency $\epsilon_{\rm rel,\;sig}$ at different energy points. Solid points represent energy points included in the training set, while hollow points represent extrapolated energy points. The gray line in (a) shows the combined efficiency of the conventional method, for which the background level in the inclusive MC sample is comparable to that of ResoSeg.}
\end{figure}

\textbf{Efficiency and generalization to other energy points.} The model is trained at three energy points, 4.22, 4.26, and 4.39\,GeV. To test its generalization, we generate about 50 thousand events at each of 4.190, 4.245, 4.360, 4.420, and 4.600\,GeV as test sets and evaluate the reconstruction efficiency on them. For comparison, we compute at the same energy points the combined efficiency $\epsilon_{\rm combined}=\sum_i \mathcal{B}_i\times \epsilon_i$ over the 16 decay channels of the conventional method, keeping the background levels in the inclusive MC samples comparable. Fig.~\ref{fig:ecms_default}(a) shows the combined efficiency of ResoSeg together with that of the conventional method. ResoSeg clearly achieves a significantly higher reconstruction efficiency; averaged over these energy points, its combined efficiency is about 2.6 times that of the conventional value. Fig.~\ref{fig:ecms_default}(b) shows the corresponding $\epsilon_{\rm rel,\;sig}$, which remains high on energy points outside the training set but degrades noticeably as the center-of-mass energy increases.

\subsection{Transfer learning}
\label{Sec:fine-tuning}

\begin{figure}
	\begin{subfigure}[b]{0.45\columnwidth}
		%\centering
		\includegraphics[width=\textwidth]{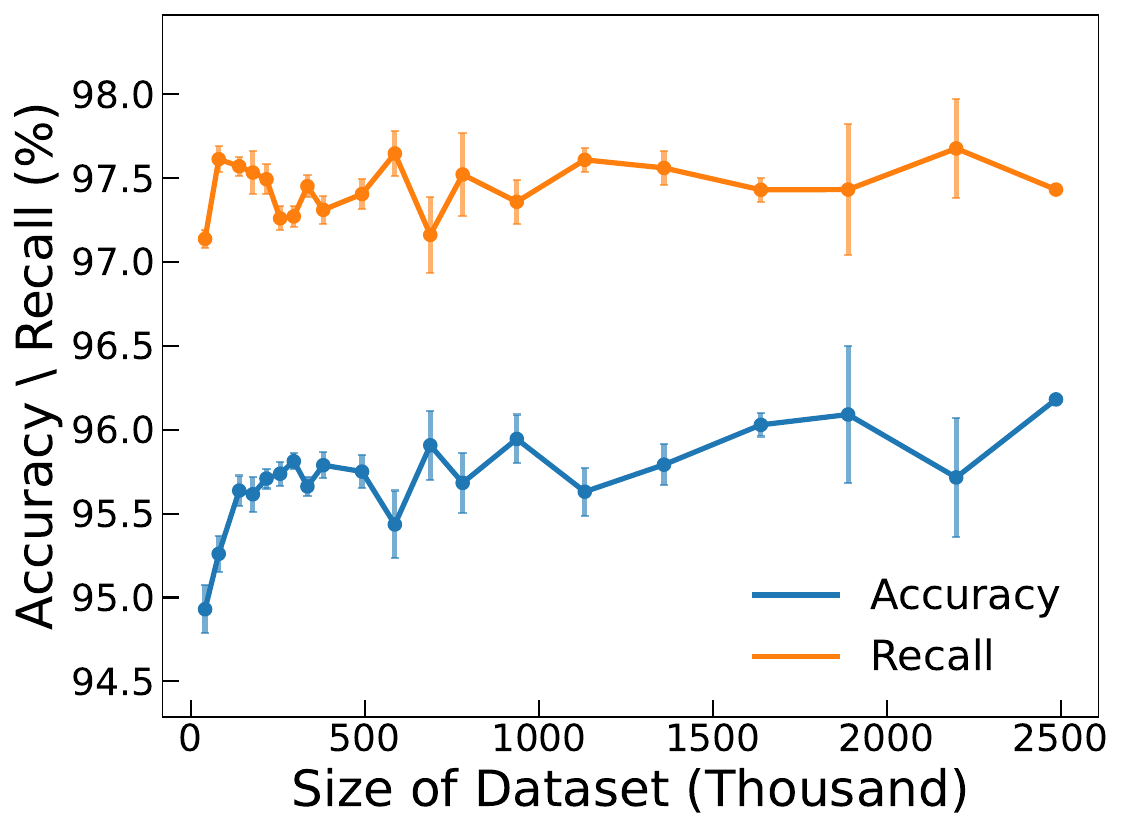}
		\caption{Accuracy and recall}
		\label{fig:ft_sub1}
	\end{subfigure}
	\begin{subfigure}[b]{0.45\columnwidth}
		%\centering
		\includegraphics[width=\textwidth]{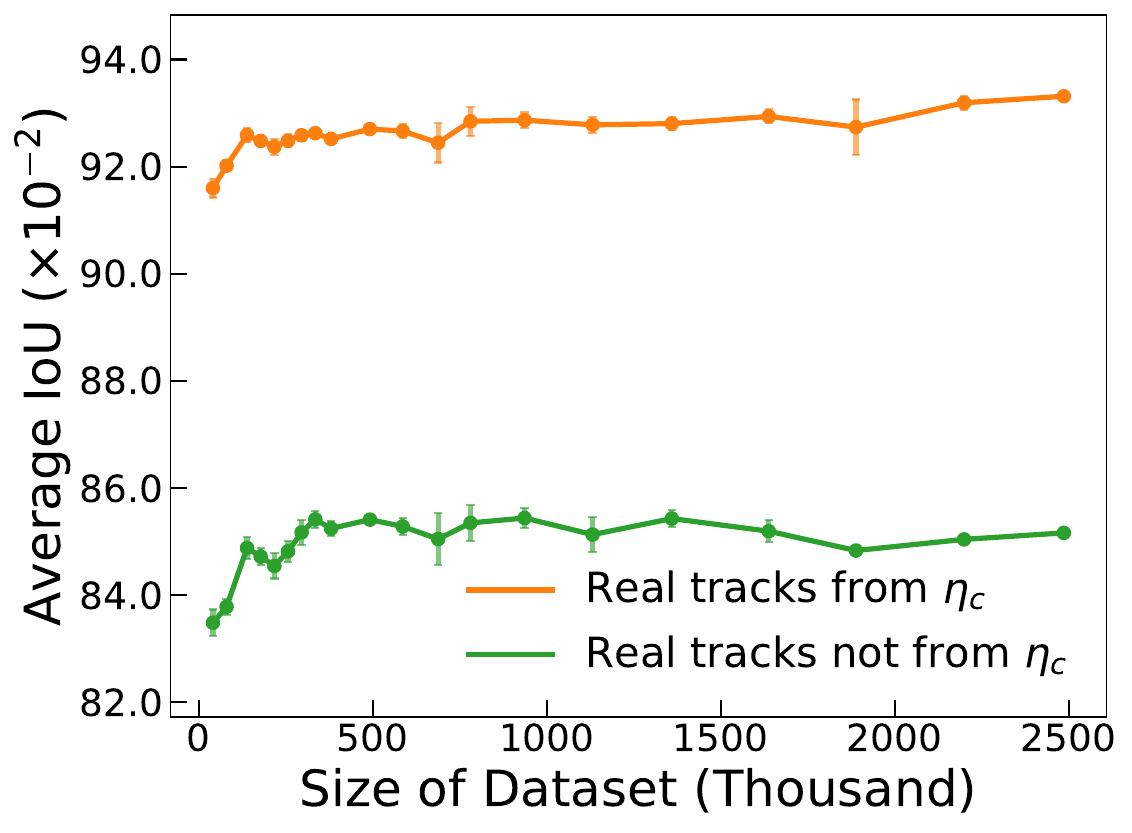}
		\caption{Average IoU score}
		\label{fig:ft_sub2}
	\end{subfigure}\\
	\begin{subfigure}[b]{0.45\columnwidth}
		%\centering
		\includegraphics[width=\textwidth]{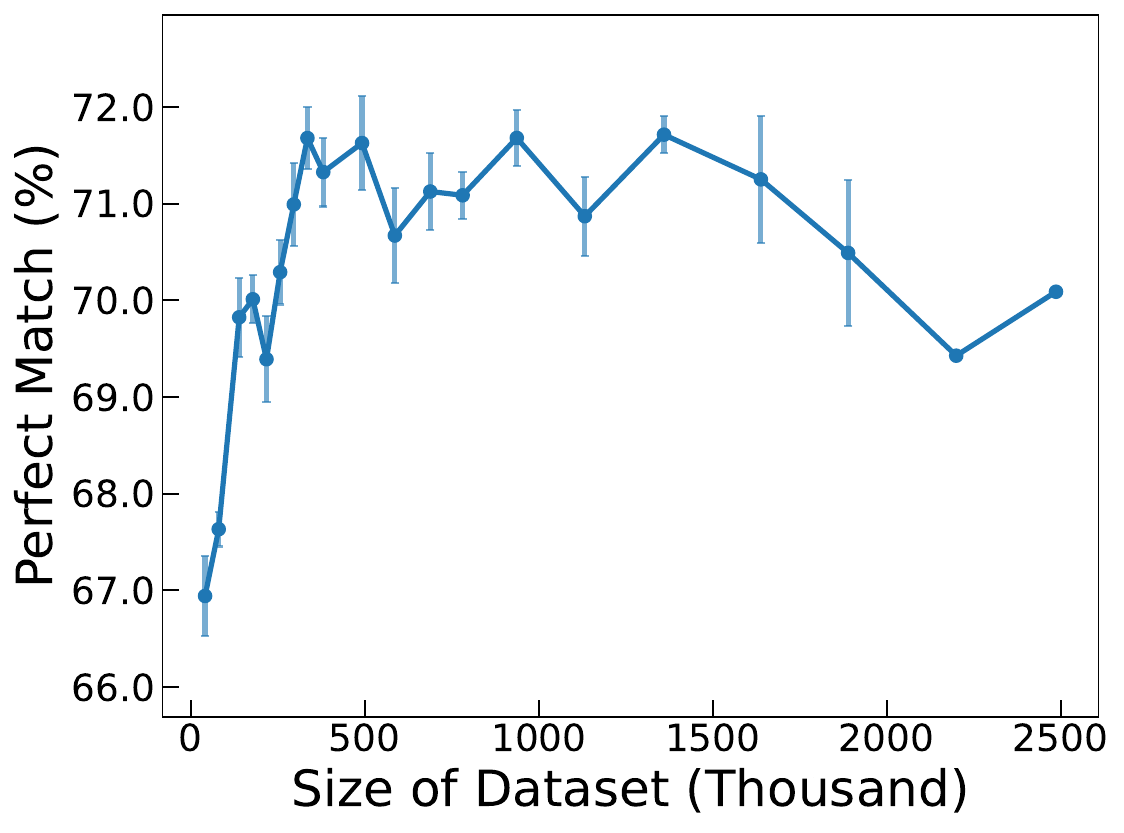}
		\caption{PMR}
		\label{fig:ft_sub3}
	\end{subfigure}
	\begin{subfigure}[b]{0.45\columnwidth}
		%\centering
		\includegraphics[width=\textwidth]{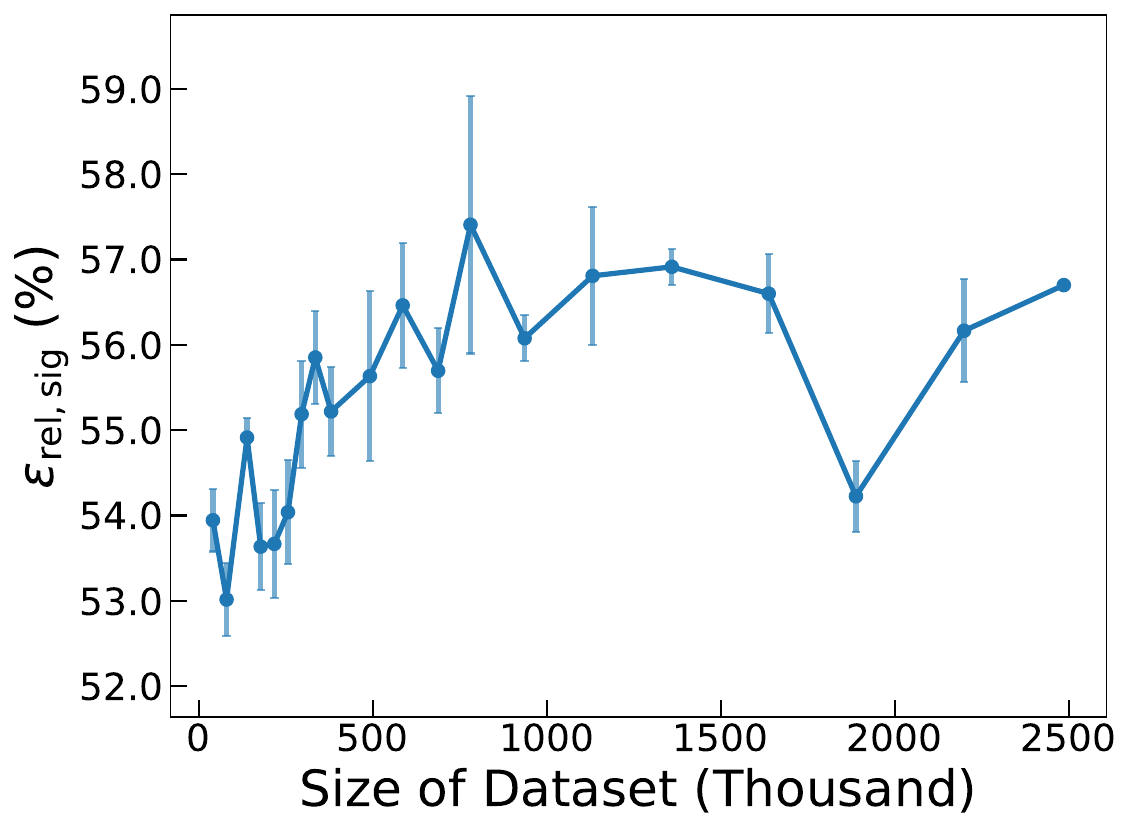}
		\caption{$\epsilon_{\rm rel,\;sig}$}
		\label{fig:ft_sub4}
	\end{subfigure}
	\caption{\raggedright \label{fig:finetune_result} The evaluation metrics for transfer learning with different training set sizes. Each data point is obtained from multiple independent training runs, and the error bars represent the SEM of these runs.}
\end{figure}

%Since the model with GA achieves the best overall performance, we choose to fine-tune this model to test its generalization capability to other decay channels. 
The dataset used for transfer learning is essentially the same as that described in Sec.~\ref{Sec:dataset}, except that the signal channel is changed to $\psi(2S)\to\gamma\eta_c(2S)$,\;$\eta_c(2S)\to\gamma h_c$,\;$h_c\to\gamma\eta_c$,\;$\eta_c\to\text{anything}$, the background channel to $\psi(2S)\to\gamma\eta_c(2S)$,\;$\eta_c(2S)\to\gamma h_c$,\;$h_c\to\text{anything}$, and the center-of-mass energy to 3.686\,GeV. The transfer learning dataset is also split into training, validation, and test sets in a ratio of 8:1:1. The post-processing procedure for $\psi(2S)\to\gamma\eta_c(2S)$,\;$\eta_c(2S)\to\gamma h_c$,\;$h_c\to\gamma\eta_c$,\;$\eta_c\to\text{anything}$ is slightly different from that for $e^+e^-\to\pi^+\pi^- h_c$,\;$h_c\to \gamma\eta_c$,\;$\eta_c\to \text{anything}$. We additionally require at least one photon in each of the three energy intervals $(0,0.07]~\text{GeV}$, $[0.08,0.16]~\text{GeV}$, and $[0.2,+\infty)~\text{GeV}$, corresponding to the three photons not originating from $\eta_c$.
% use perfect match rate instead
The evaluation metrics for transfer learning with different data sizes are shown in Fig.~\ref{fig:finetune_result}; each training run uses 100 epochs. Each data point is obtained by averaging over multiple independent training runs, with the number of repetitions chosen according to the dataset size so as to keep the computational cost manageable. The statistical uncertainty of each data point is taken as the standard error of the mean (SEM) of these repeated runs and is shown as error bars in the figure.
As few as 300 thousand events already yield good performance, and the metrics saturate once the data size reaches 800 thousand.

Notably, the model's ability to distinguish tracks not originating from $\eta_c$ is markedly weaker on the transfer learning dataset than on the BESIII-$\eta_c$ dataset, leading to slightly degraded PMR and $\epsilon_{\rm rel,\;sig}$. This is physically plausible because the transfer learning dataset contains two low-energy photons among the three non-$\eta_c$ tracks, which are inherently difficult to identify correctly.

\subsection{Robustness Evaluation}
\label{Sec:robustness}
Robustness is evaluated by shifting the mass ($M$) and width ($\Gamma$) of the $\eta_c$ in simulation, as well as by altering the branching fractions of $\eta_c$ in the dataset. 
In the simulation, the default values of $\eta_c$ mass and width are $M = 2.9839\ \mathrm{GeV}/c^2$ and $\Gamma = 0.032\ \mathrm{GeV}/c^2$. Based on the uncertainties of the $\eta_c$ from the PDG~\cite{PDG}, we shift $M$ by $\pm 10\sigma$ ($0.004\ \mathrm{GeV}/c^2$) and $\pm 5\sigma$ ($0.002\ \mathrm{GeV}/c^2$), and shift $\Gamma$ by $\pm 5\sigma$ ($0.0025\ \mathrm{GeV}/c^2$) and $\pm 2.5\sigma$ ($0.00125\ \mathrm{GeV}/c^2$). 
We quantify robustness by the change in $\epsilon_{\rm rel,\;sig}$, $\Delta\epsilon_{\rm rel,\;sig}=|\epsilon_{\rm rel,\;sig,\,modified} - \epsilon_{\rm rel,\;sig,\,default}|/\epsilon_{\rm rel,\;sig,\,default}$.
%While the final efficiency change $\Delta \epsilon=|\epsilon_{\rm modified} - \epsilon_{\rm default}|/\epsilon_{\rm default}$ can be regarded as the systematic uncertainty when applying the method to real data for physical analysis.

\begin{figure}
	%\begin{subfigure}[b]{0.45\columnwidth}
	%	\centering
	%	\includegraphics[width=\textwidth]{pics/mass_shift_eff_err_abs_mean.pdf}
	%	\caption{}   % 自动生成 (a)
	%	\label{fig:sub1}
	%\end{subfigure}
	\begin{subfigure}[b]{0.45\textwidth}
		%\centering
		\includegraphics[width=\textwidth]{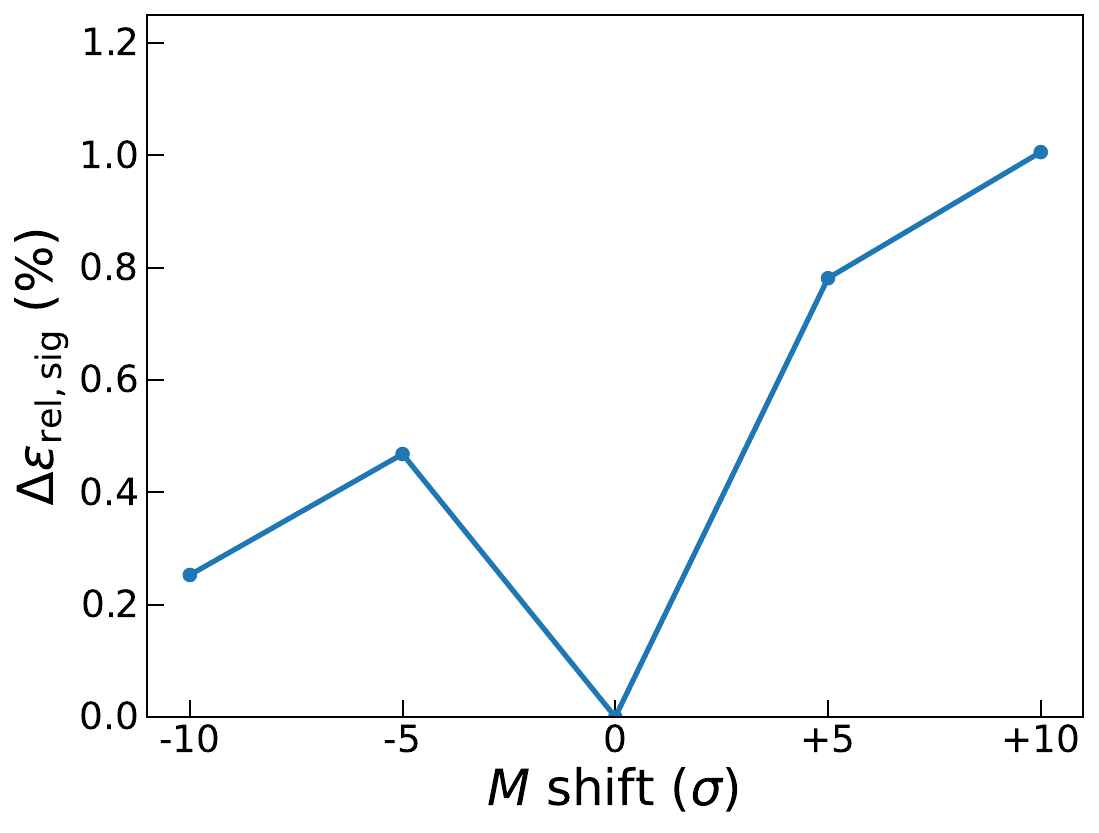}
		\caption{$\epsilon_{\rm rel,\;sig}$ against mass shift.}
		\label{fig:ms_sub1}
	\end{subfigure}\\
	%\begin{subfigure}[b]{0.45\columnwidth}
	%	\centering
	%	\includegraphics[width=\textwidth]{pics/gamma_shift_eff_err_abs_mean.pdf}
	%	\caption{}   % 自动生成 (a)
	%	\label{fig:sub1}
	%\end{subfigure}
	\begin{subfigure}[b]{0.45\textwidth}
		%\centering
		\includegraphics[width=\textwidth]{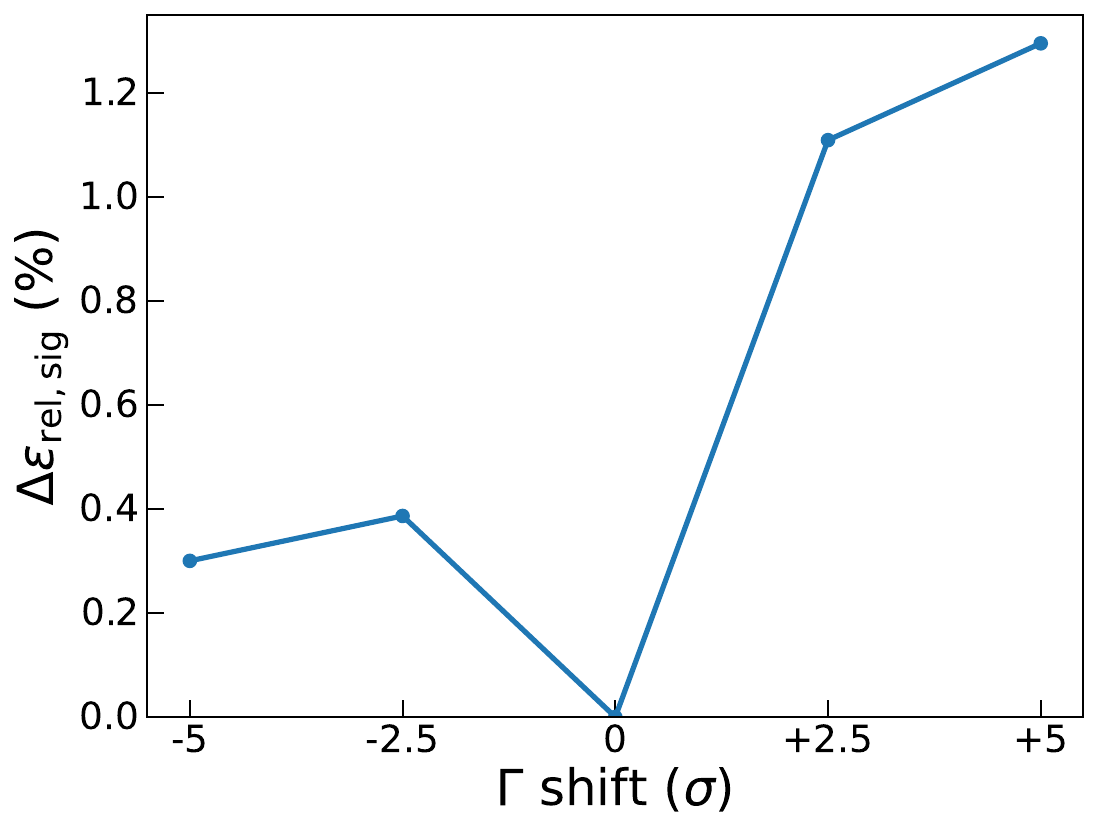}
		\caption{$\epsilon_{\rm rel,\;sig}$ against width shift.}
		\label{fig:ms_sub2}
	\end{subfigure}
	\caption{\raggedright \label{fig:ms_shift} Average $\Delta\epsilon_{\rm rel,\;sig}$ induced by (a) mass and (b) width shifts across the three energy points.}
\end{figure}

Fig.~\ref{fig:ms_shift} shows the average effect of mass shifts (a) and width shifts (b) over the three energy points. ResoSeg is robust to both mass and width shifts.
%and the associated systematic uncertainties in realistic physics analyses are also negligible.
Regarding branching fractions, the default configuration uses PDG-adjusted values. To test robustness, we change them to the BESIII default fractions and to uniform fractions.
The average $\Delta\epsilon_{\rm rel,\;sig}$ over the three energy points for the BESIII default fractions is 1.98\%, and that for the uniform fractions is 4.92\%,
indicating that the model is quite robust.

%However, it is worth noting that when the fraction changes stay within the range of real experimental scenarios (i.e., the BESIII default setting), the variation in the final efficiency $\Delta \epsilon$ is appreciable. This suggests that the systematic uncertainty induced by fraction variations is a non-negligible factor when applying the Resonant Tagger to real physics analyses.
%\begin{table}
%	\caption{\label{tab:br_shift}Average effect of fraction variations across the three energy points (\%). The small change in $\epsilon_{\rm rel}$ reflects the good robustness of the Resonant Tagger, whereas the non-negligible $\Delta \epsilon$ observed for the BESIII default setting emphasizes that systematic uncertainties due to fraction variations deserve careful attention in realistic physics analyses.}
%	\begin{ruledtabular}
%		\begin{tabular}{lcc}
%			% Lines of table here ending with \\
%			&					$\Delta \epsilon$&		$\Delta \epsilon_{\rm rel}$\\\hline
%			BESIII Default&		9.66&					2.57\\
%			Uniform&			56.84&					3.11
%
%		\end{tabular}
%	\end{ruledtabular}
%\end{table}

\section{Bayesian Optimization of the Interaction Features}
\label{Sec:bayes}
Since the underlying physics of jets and quarkonium differs considerably, we design five additional features specifically for track-pair descriptions. Together with the four features adopted in ParT~\cite{ParT}, we obtain a total of nine interaction features, as listed in Eq.~(\ref{9feat}):
\begin{equation}
	\label{9feat}
	\begin{split}
		&\Delta = \sqrt{(y_a-y_b)^2+(\phi_a-\phi_b)^2},\\
		&k_t = min(P_{T,a},P_{T,b})\Delta,\\
		&z = min(P_{T,a},P_{T,b})/ (P_{T,a}+P_{T,b}),\\
		&M^2=(E_a+E_b)^2-|\vec{P_a}+\vec{P_b}|^2,\\
		&RM^2=(E_{\rm cms}-E_a-E_b)^2-|\vec{P_{\rm cms}}-\vec{P_a}-\vec{P_b}|^2,\\
		&Q = q_a*q_b,\\
		&|{\rm Helicity\;Angle}| = \left|\frac{\vec{P^*_b}\cdot\hat{z}}  {|\vec{P^*_b}|}\right|,\\
		&{\rm Polar\;Angle} = \frac{\vec{P_{a,z}}+\vec{P_{b,z}}}  {|\vec{P_a}+\vec{P_b}|},\\
		&|\Delta E_{\rm EMC}| = |E_{a,{\rm EMC}}-E_{b,{\rm EMC}}|,
	\end{split}
\end{equation}
where $\vec{P_i} =(P_{x,i}, P_{y,i}, P_{z,i})$ is the three-momentum in the laboratory frame, and $\vec{P}_i^*$ is that in the center-of-mass frame of the two-particle system, $P_{T,i}=(P_{x,i}^2+P_{y,i}^2)^{1/2}$ is the transverse momentum in the laboratory frame, and $E_{i,{\rm EMC}}$ is the deposit energy in EMC, for $i = a,\;b$.

To efficiently search the combinatorial feature space, we adopt Bayesian optimization, specifically the Efficient Global Optimization framework. We construct a Gaussian process surrogate model to provide predictive mean
% $\hat{y}(x)$ 
and variance 
%$s^2(x)$
for any feature subset 
%$x$
. At each iteration, the next subset is selected by maximizing the Expected Improvement acquisition function,
%:
%\begin{multline}
%	EI(x) = (y_{\min} - \hat{y}(x))\,\Phi\!\left(\frac{y_{\min} - \hat{y}(x)}{s(x)}\right)\\ + s(x)\,\phi\!\left(\frac{y_{\min} - \hat{y}(x)}{s(x)}\right),
%\end{multline}
%where $y_{\min}$ is the current best observed value (assuming minimization of the negative validation performance), and $\Phi$ and $\phi$ are the cumulative distribution function and probability density function of the standard normal distribution, respectively.
This procedure balances exploration and exploitation with a limited number of costly model retraining evaluations~\cite{bayes}.

The optimal subset of interaction features is found to be $[M^2,\; RM^2,\; Q,\; k_t]$. Table~\ref{tab:feature_comp} compares the performance of this optimal subset with that of the default ParT interaction subset, when applied to both ResoSeg and ParT. It is evident that, for quarkonium, the optimized interaction features can significantly improve both $\epsilon_{\rm rel,\;sig}$ and ${\rm Rej}_{90}$ for ResoSeg, and also enhance ${\rm Rej}_{90}$ for ParT.

\begin{table*}
	\caption{\raggedright \label{tab:feature_comp} Evaluation metrics of ResoSeg and ParT using the optimal subset $[M^2,\; RM^2,\; Q,\; k_t]$ and the default ParT subset $[M^2,\Delta,k_t,z]$ on the test set. For the average IoU, results are presented as ``real tracks from $\eta_c$'' / ``real tracks not from $\eta_c$''. The $\epsilon_{\rm rel,\;sig}$ is the signal efficiency obtained after imposing the $\eta_c$ and $h_c$ mass-window selections together with the event score requirement, while the ${\rm Rej}_{90}$ is evaluated using only the event score selection, at the event score threshold that yields a relative signal efficiency of 90\%.}
	\begin{ruledtabular}
		\begin{tabular}{lcccccc}
			&					Acc(\%)&	Recall(\%)&	Average IoU($\times 10^{-2}$)&	PMR(\%)&	$\epsilon_{\rm rel,\;sig}$(\%)&	${\rm Rej}_{90}$\\\hline
			Optimal ResoSeg&	97.31&		98.84&		96.12/95.20&					87.91&		79.79&							47.82\\
			Default ResoSeg&	97.14&		98.70&		94.68/92.61&					82.26&		74.81&							15.94\\
			Optimal ParT&		95.93&		98.93&		\textbackslash&		\textbackslash&		\textbackslash&						19.69\\
			Default ParT&		96.16&		99.16&		\textbackslash&		\textbackslash&		\textbackslash&						17.02
		\end{tabular}
	\end{ruledtabular}
\end{table*}

\begin{figure}
	\includegraphics[width=0.9\columnwidth]{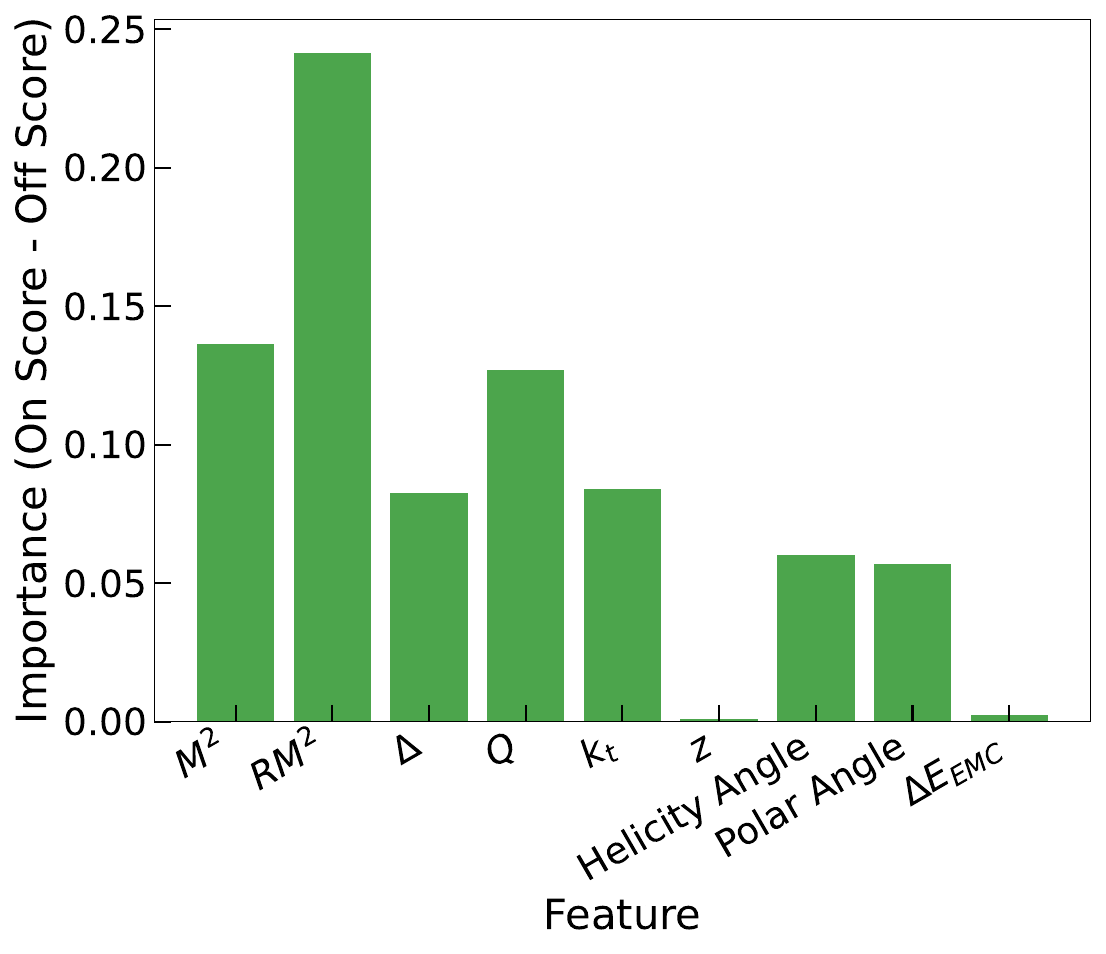}
	\caption{\raggedright \label{fig:importance}The importance of each interaction feature, quantified by the average change in $\epsilon_{\rm rel,\;sig}$ when the feature is included versus excluded.}
\end{figure}

To quantify the individual contribution of each feature, we define its importance as the average change in $\epsilon_{\rm rel,\;sig}$ when the feature is included versus when it is excluded, averaged over all evaluated subsets. The resulting importances are presented in Fig.~\ref{fig:importance}. $M^2$ and $RM^2$ yield the most substantial improvements to the model, while the other features contribute only marginally in comparison.

\section{Gated attention and conditional layer normalization}
\label{Sec:gate}
Based on the aforementioned model as a baseline, we introduce three model modifications: (1) replacing all attention layers in the model with gated attention (GA)~\cite{gated_attn}; (2) replacing all layer normalization with conditional layer normalization (CLN)~\cite{CLN}, where the condition consists of the number of charged tracks with EMC information, the number of charged tracks without EMC information, the number of neutral tracks, and $E_{\rm cms}$; (3) applying both (1) and (2). The corresponding evaluation metrics are summarized in Table~\ref{tab:metrics}. 

Interestingly, when the interaction features exclude $RM^2$ and include only $[M^2, Q, k_t]$, the model with CLN (denoted as noRM\_CLN) outperforms the one without CLN (denoted as noRM) by a 6.61\% improvement in PMR, a 5.87\% gain in $\epsilon_{\rm rel,\;sig}$, and a 44.68\% improvement in ${\rm Rej}_{90}$. However, this improvement disappears when $RM^2$ is included in the interaction features, likely because $RM^2$ already contains the event-level information $E_{\rm cms}$, which serves a similar role to the condition used in CLN.

Given that no significant difference is observed among these models, we opt not to introduce GA and CLN in the final model.
\begin{table*}
	\caption{\raggedright \label{tab:metrics}Evaluation metrics of the baseline and modified models on the test set. For the average IoU, results are presented as ``real tracks from $\eta_c$'' / ``real tracks not from $\eta_c$''. The $\epsilon_{\rm rel,\;sig}$ is the signal efficiency obtained after imposing the $\eta_c$ and $h_c$ mass-window selections together with the event score requirement, while the ${\rm Rej}_{90}$ is evaluated using only the event score selection, at the event score threshold that yields a relative signal efficiency of 90\%.}
	\begin{ruledtabular}
		\begin{tabular}{lcccccc}
			&					Acc(\%)&	Recall(\%)&	Average IoU($\times 10^{-2}$)&	PMR(\%)&	$\epsilon_{\rm rel,\;sig}$(\%)&	${\rm Rej}_{90}$\\\hline
			Baseline&			97.31&		98.84&		96.12/95.20&					87.91&		79.79&						47.82\\
			GA&					97.23&		98.84&		96.00/95.01&					88.08&		80.30&						46.46\\
			CLN&				97.50&		98.97&		96.23/95.23&					88.17&		80.52&						42.18\\
			GA + CLN&			97.11&		99.09&		95.90/95.01&					87.23&		80.56&						34.71\\
			noRM&				96.71&		99.08&		94.78/92.73&					82.21&		76.10&						26.77\\
			noRM + CLN&			97.42&		99.01&		96.07/94.92&					87.64&		80.57&						38.73
		\end{tabular}
	\end{ruledtabular}
\end{table*}

%\begin{figure}
%	\includegraphics[width=0.9\columnwidth]{pics/ecms_diff_ver_rel.pdf}
%	\caption{\raggedright \label{fig:eff_diff}The ResoSeg efficiency $\epsilon_{\rm rel}$ of the baseline and modified models at different energy points.}
%\end{figure}
%\section{Discussion}
%In this section, we present several attempts to improve the model performance. In Sec.~\ref{Sec:bayes}, we test different combinations of interaction features and identify the optimal set via Bayesian optimization~\cite{bayes}. In Sec.~\ref{Sec:gate}, we investigate how the introduction of gated attention (GA)~\cite{gated_attn} and conditional layer normalization (CLN)~\cite{CLN} affects the model performance.

\section{Conclusion and Discussion}

In this work, we integrate segmentation into experimental high-energy physics analyses and, for the first time, propose a model that simultaneously performs particle-level segmentation and event-level classification, thereby enabling the unified reconstruction of intermediate resonances in a single pass. To this end, we construct a dedicated BESIII-$\eta_c$ dataset for the process $e^+e^-\to\pi^+\pi^- h_c$, $h_c\to\gamma\eta_c$, $\eta_c\to\text{anything}$, and propose ResoSeg, which combines ParT's particle-wise attention~\cite{ParT} with a SAM-inspired mask decoder~\cite{SAM}.

At the event level, ResoSeg is comparable to ParT~\cite{ParT}, and gives stronger background rejection on the more complex inclusive MC sample. By reconstructing the full four-momentum of the intermediate resonances, it provides further background suppression: at comparable background levels, the average combined efficiency of ResoSeg is 2.6 times that of the conventional method. These improvements enable more precise measurements of cross sections, masses, widths, and quantum numbers of exotic candidates such as the $Z_c$ states. In the inclusive MC sample, the model also separates signal from background stably, as verified by repeated input/output checks.

ResoSeg also shows good generalization and robustness: it remains effective at extrapolated energy points from 4.19 to 4.60~GeV, adapts to an alternative $\eta_c$ production chain via transfer learning, and is stable under shifts in the $\eta_c$ mass, width, and fractions.

Although demonstrated for $\eta_c$ at BESIII, the approach is inherently general and can be transferred to other resonances and experiments, especially other lepton collider experiments. Future work will extend the model to additional resonances, address the data imbalance among the many $\eta_c$ decay modes, and further optimize the architecture. Interference between resonance states, conventionally handled through partial-wave analysis, may instead be captured directly by deep learning in latent space, though a systematic treatment remains to be developed. Beyond resonance tagging, per-track semantic labels may also serve emerging paradigms such as foundation models for particle physics~\cite{evenet,omniLearned,omnimol} and simulation-based inference~\cite{Inference1,Inference2,Inference3}.

\section{Acknowledgement}
This project is supported by the Strategic Priority Research Program of Chinese Academy of Sciences under Grant XDA0480600,
%junhao
the National Natural Science Foundation of China and Research Grants No. 12405099, 
and the Fundamental Research Funds for the Central Universities under Contract No. 010-63263110. 

The authors thank the BESIII Collaboration for providing the simulation framework and 
the IHEP Computing Center for their strong support. 
The authors acknowledge the use of DeepSeek V4 Pro for code debugging. 

\bibliography{bibitem}

\begin{thebibliography}{38}%
\makeatletter
\providecommand \@ifxundefined [1]{%
 \@ifx{#1\undefined}
}%
\providecommand \@ifnum [1]{%
 \ifnum #1\expandafter \@firstoftwo
 \else \expandafter \@secondoftwo
 \fi
}%
\providecommand \@ifx [1]{%
 \ifx #1\expandafter \@firstoftwo
 \else \expandafter \@secondoftwo
 \fi
}%
\providecommand \natexlab [1]{#1}%
\providecommand \enquote  [1]{``#1''}%
\providecommand \bibnamefont  [1]{#1}%
\providecommand \bibfnamefont [1]{#1}%
\providecommand \citenamefont [1]{#1}%
\providecommand \href@noop [0]{\@secondoftwo}%
\providecommand \href [0]{\begingroup \@sanitize@url \@href}%
\providecommand \@href[1]{\@@startlink{#1}\@@href}%
\providecommand \@@href[1]{\endgroup#1\@@endlink}%
\providecommand \@sanitize@url [0]{\catcode `\\12\catcode `\$12\catcode
  `\&12\catcode `\#12\catcode `\^12\catcode `\_12\catcode `\%12\relax}%
\providecommand \@@startlink[1]{}%
\providecommand \@@endlink[0]{}%
\providecommand \url  [0]{\begingroup\@sanitize@url \@url }%
\providecommand \@url [1]{\endgroup\@href {#1}{\urlprefix }}%
\providecommand \urlprefix  [0]{URL }%
\providecommand \Eprint [0]{\href }%
\providecommand \doibase [0]{https://doi.org/}%
\providecommand \selectlanguage [0]{\@gobble}%
\providecommand \bibinfo  [0]{\@secondoftwo}%
\providecommand \bibfield  [0]{\@secondoftwo}%
\providecommand \translation [1]{[#1]}%
\providecommand \BibitemOpen [0]{}%
\providecommand \bibitemStop [0]{}%
\providecommand \bibitemNoStop [0]{.\EOS\space}%
\providecommand \EOS [0]{\spacefactor3000\relax}%
\providecommand \BibitemShut  [1]{\csname bibitem#1\endcsname}%
\let\auto@bib@innerbib\@empty
%</preamble>
\bibitem [{\citenamefont {Brambilla}\ \emph {et~al.}(2020)\citenamefont
  {Brambilla}, \citenamefont {Eidelman}, \citenamefont {Hanhart}, \citenamefont
  {Nefediev}, \citenamefont {Shen}, \citenamefont {Thomas}, \citenamefont
  {Vairo},\ and\ \citenamefont {Yuan}}]{Brambilla:2019esw}%
  \BibitemOpen
  \bibfield  {author} {\bibinfo {author} {\bibfnamefont {N.}~\bibnamefont
  {Brambilla}}, \bibinfo {author} {\bibfnamefont {S.}~\bibnamefont {Eidelman}},
  \bibinfo {author} {\bibfnamefont {C.}~\bibnamefont {Hanhart}}, \bibinfo
  {author} {\bibfnamefont {A.}~\bibnamefont {Nefediev}}, \bibinfo {author}
  {\bibfnamefont {C.-P.}\ \bibnamefont {Shen}}, \bibinfo {author}
  {\bibfnamefont {C.~E.}\ \bibnamefont {Thomas}}, \bibinfo {author}
  {\bibfnamefont {A.}~\bibnamefont {Vairo}},\ and\ \bibinfo {author}
  {\bibfnamefont {C.-Z.}\ \bibnamefont {Yuan}},\ }\bibfield  {title} {\bibinfo
  {title} {{The $XYZ$ states: experimental and theoretical status and
  perspectives}},\ }\href {https://doi.org/10.1016/j.physrep.2020.05.001}
  {\bibfield  {journal} {\bibinfo  {journal} {Phys. Rept.}\ }\textbf {\bibinfo
  {volume} {873}},\ \bibinfo {pages} {1} (\bibinfo {year} {2020})},\ \Eprint
  {https://arxiv.org/abs/1907.07583} {arXiv:1907.07583 [hep-ex]} \BibitemShut
  {NoStop}%
\bibitem [{\citenamefont {Qu}\ \emph {et~al.}(2024)\citenamefont {Qu},
  \citenamefont {Li},\ and\ \citenamefont {Qian}}]{ParT}%
  \BibitemOpen
  \bibfield  {author} {\bibinfo {author} {\bibfnamefont {H.}~\bibnamefont
  {Qu}}, \bibinfo {author} {\bibfnamefont {C.}~\bibnamefont {Li}},\ and\
  \bibinfo {author} {\bibfnamefont {S.}~\bibnamefont {Qian}},\ }\href
  {https://arxiv.org/abs/2202.03772} {\bibinfo {title} {Particle transformer
  for jet tagging}} (\bibinfo {year} {2024}),\ \Eprint
  {https://arxiv.org/abs/2202.03772} {arXiv:2202.03772 [hep-ph]} \BibitemShut
  {NoStop}%
\bibitem [{\citenamefont {Shmakov}\ \emph {et~al.}(2022)\citenamefont
  {Shmakov}, \citenamefont {Fenton}, \citenamefont {Ho}, \citenamefont {Hsu},
  \citenamefont {Whiteson},\ and\ \citenamefont {Baldi}}]{SPANet}%
  \BibitemOpen
  \bibfield  {author} {\bibinfo {author} {\bibfnamefont {A.}~\bibnamefont
  {Shmakov}}, \bibinfo {author} {\bibfnamefont {M.~J.}\ \bibnamefont {Fenton}},
  \bibinfo {author} {\bibfnamefont {T.-W.}\ \bibnamefont {Ho}}, \bibinfo
  {author} {\bibfnamefont {S.-C.}\ \bibnamefont {Hsu}}, \bibinfo {author}
  {\bibfnamefont {D.}~\bibnamefont {Whiteson}},\ and\ \bibinfo {author}
  {\bibfnamefont {P.}~\bibnamefont {Baldi}},\ }\bibfield  {title} {\bibinfo
  {title} {Spanet: Generalized permutationless set assignment for particle
  physics using symmetry preserving attention},\ }\bibfield  {journal}
  {\bibinfo  {journal} {SciPost Physics}\ }\textbf {\bibinfo {volume} {12}},\
  \href {https://doi.org/10.21468/scipostphys.12.5.178}
  {10.21468/scipostphys.12.5.178} (\bibinfo {year} {2022})\BibitemShut
  {NoStop}%
\bibitem [{GN3(2026)}]{GN3}%
  \BibitemOpen
  \bibfield  {title} {\bibinfo {title} {{GN3: Multi-task, Multi-modal
  Transformers for Jet Flavour Tagging in ATLAS}},\ }\href@noop {} {\
  (\bibinfo {year} {2026})}\BibitemShut {NoStop}%
\bibitem [{\citenamefont {Ablikim}\ \emph
  {et~al.}(2025{\natexlab{a}})\citenamefont {Ablikim} \emph
  {et~al.}}]{ParT_work1}%
  \BibitemOpen
  \bibfield  {author} {\bibinfo {author} {\bibfnamefont {M.}~\bibnamefont
  {Ablikim}} \emph {et~al.} (\bibinfo {collaboration} {BESIII Collaboration}),\
  }\bibfield  {title} {\bibinfo {title} {Observation of the singly
  cabibbo-suppressed decay
  ${\mathrm{\ensuremath{\Lambda}}}_{c}^{+}\ensuremath{\rightarrow}p{\ensuremath{\pi}}^{0}$},\
  }\href {https://doi.org/10.1103/PhysRevD.111.L051101} {\bibfield  {journal}
  {\bibinfo  {journal} {Phys. Rev. D}\ }\textbf {\bibinfo {volume} {111}},\
  \bibinfo {pages} {L051101} (\bibinfo {year}
  {2025}{\natexlab{a}})}\BibitemShut {NoStop}%
\bibitem [{\citenamefont {Ablikim}\ \emph {et~al.}(2026)\citenamefont {Ablikim}
  \emph {et~al.}}]{ParT_work2}%
  \BibitemOpen
  \bibfield  {author} {\bibinfo {author} {\bibfnamefont {M.}~\bibnamefont
  {Ablikim}} \emph {et~al.} (\bibinfo {collaboration} {BESIII Collaboration}),\
  }\href {https://arxiv.org/abs/2602.11974} {\bibinfo {title} {Measurement of
  the singly cabibbo-suppressed decay $\lambda_c^+\to p\eta'$ with deep
  learning}} (\bibinfo {year} {2026}),\ \Eprint
  {https://arxiv.org/abs/2602.11974} {arXiv:2602.11974 [hep-ex]} \BibitemShut
  {NoStop}%
\bibitem [{\citenamefont {Ablikim}\ \emph
  {et~al.}(2025{\natexlab{b}})\citenamefont {Ablikim} \emph
  {et~al.}}]{ParT_work3}%
  \BibitemOpen
  \bibfield  {author} {\bibinfo {author} {\bibfnamefont {M.}~\bibnamefont
  {Ablikim}} \emph {et~al.} (\bibinfo {collaboration} {BESIII Collaboration}),\
  }\bibfield  {title} {\bibinfo {title} {Search for radiative leptonic decay
  ${D^+}\to\gamma e^+\nu_e$ using deep learning*},\ }\href
  {https://doi.org/10.1088/1674-1137/adcdf3} {\bibfield  {journal} {\bibinfo
  {journal} {Chinese Physics C}\ }\textbf {\bibinfo {volume} {49}},\ \bibinfo
  {pages} {083001} (\bibinfo {year} {2025}{\natexlab{b}})}\BibitemShut
  {NoStop}%
\bibitem [{\citenamefont {Navas}\ \emph {et~al.}(2024)\citenamefont {Navas}
  \emph {et~al.}}]{PDG}%
  \BibitemOpen
  \bibfield  {author} {\bibinfo {author} {\bibfnamefont {S.}~\bibnamefont
  {Navas}} \emph {et~al.} (\bibinfo {collaboration} {Particle Data Group
  Collaboration}),\ }\bibfield  {title} {\bibinfo {title} {Review of particle
  physics},\ }\href {https://doi.org/10.1103/PhysRevD.110.030001} {\bibfield
  {journal} {\bibinfo  {journal} {Phys. Rev. D}\ }\textbf {\bibinfo {volume}
  {110}},\ \bibinfo {pages} {030001} (\bibinfo {year} {2024})}\BibitemShut
  {NoStop}%
\bibitem [{\citenamefont {Ablikim}\ \emph
  {et~al.}(2017{\natexlab{a}})\citenamefont {Ablikim} \emph
  {et~al.}}]{traditional}%
  \BibitemOpen
  \bibfield  {author} {\bibinfo {author} {\bibfnamefont {M.}~\bibnamefont
  {Ablikim}} \emph {et~al.} (\bibinfo {collaboration} {BESIII Collaboration}),\
  }\bibfield  {title} {\bibinfo {title} {Evidence of two resonant structures in
  $e^+e^-\to\pi^+\pi^-h_c$},\ }\bibfield  {journal} {\bibinfo  {journal}
  {Physical Review Letters}\ }\textbf {\bibinfo {volume} {118}},\ \href
  {https://doi.org/10.1103/physrevlett.118.092002}
  {10.1103/physrevlett.118.092002} (\bibinfo {year}
  {2017}{\natexlab{a}})\BibitemShut {NoStop}%
\bibitem [{\citenamefont {Ablikim}\ \emph
  {et~al.}(2025{\natexlab{c}})\citenamefont {Ablikim} \emph
  {et~al.}}]{traditional2}%
  \BibitemOpen
  \bibfield  {author} {\bibinfo {author} {\bibfnamefont {M.}~\bibnamefont
  {Ablikim}} \emph {et~al.} (\bibinfo {collaboration} {BESIII Collaboration}),\
  }\bibfield  {title} {\bibinfo {title} {Search for ${1}^{\ensuremath{-}+}$
  charmoniumlike hybrid via
  ${e}^{+}{e}^{\ensuremath{-}}\ensuremath{\rightarrow}\ensuremath{\gamma}{\ensuremath{\eta}}^{(\ensuremath{'})}{\ensuremath{\eta}}_{c}$
  at center-of-mass energies between 4.258 and 4.681 gev},\ }\href
  {https://doi.org/10.1103/2rq2-nr4m} {\bibfield  {journal} {\bibinfo
  {journal} {Phys. Rev. D}\ }\textbf {\bibinfo {volume} {111}},\ \bibinfo
  {pages} {112007} (\bibinfo {year} {2025}{\natexlab{c}})}\BibitemShut
  {NoStop}%
\bibitem [{\citenamefont {Ablikim}\ \emph {et~al.}(2021)\citenamefont {Ablikim}
  \emph {et~al.}}]{traditional3}%
  \BibitemOpen
  \bibfield  {author} {\bibinfo {author} {\bibfnamefont {M.}~\bibnamefont
  {Ablikim}} \emph {et~al.} (\bibinfo {collaboration} {BESIII Collaboration}),\
  }\bibfield  {title} {\bibinfo {title} {Measurements of
  ${e}^{+}{e}^{\ensuremath{-}}\ensuremath{\rightarrow}{\ensuremath{\eta}}_{\mathrm{c}}{\ensuremath{\pi}}^{+}{\ensuremath{\pi}}^{\ensuremath{-}}{\ensuremath{\pi}}^{0}$,
  ${\ensuremath{\eta}}_{\mathrm{c}}{\ensuremath{\pi}}^{+}{\ensuremath{\pi}}^{\ensuremath{-}}$,
  and
  ${\ensuremath{\eta}}_{\mathrm{c}}{\ensuremath{\pi}}^{0}\ensuremath{\gamma}$
  at $\sqrt{s}$ from 4.18 to 4.60 gev, and search for a ${Z}_{\mathrm{c}}$
  state close to the $d\overline{D}$ threshold decaying to
  ${\ensuremath{\eta}}_{\mathrm{c}}\ensuremath{\pi}$ at $\sqrt{s}=4.23\text{
  }\text{ }\mathrm{GeV}$},\ }\href
  {https://doi.org/10.1103/PhysRevD.103.032006} {\bibfield  {journal} {\bibinfo
   {journal} {Phys. Rev. D}\ }\textbf {\bibinfo {volume} {103}},\ \bibinfo
  {pages} {032006} (\bibinfo {year} {2021})}\BibitemShut {NoStop}%
\bibitem [{\citenamefont {Ablikim}\ \emph {et~al.}(2010)\citenamefont {Ablikim}
  \emph {et~al.}}]{Ablikim:2009aa}%
  \BibitemOpen
  \bibfield  {author} {\bibinfo {author} {\bibfnamefont {M.}~\bibnamefont
  {Ablikim}} \emph {et~al.} (\bibinfo {collaboration} {BESIII Collaboration}),\
  }\bibfield  {title} {\bibinfo {title} {Design and construction of the besiii
  detector},\ }\href@noop {} {\bibfield  {journal} {\bibinfo  {journal} {Nucl.
  Instrum. Meth. A}\ }\textbf {\bibinfo {volume} {614}},\ \bibinfo {pages}
  {345} (\bibinfo {year} {2010})}\BibitemShut {NoStop}%
\bibitem [{\citenamefont {Yu}\ \emph {et~al.}(2016)\citenamefont {Yu} \emph
  {et~al.}}]{Yu:IPAC2016-TUYA01}%
  \BibitemOpen
  \bibfield  {author} {\bibinfo {author} {\bibfnamefont {C.~H.}\ \bibnamefont
  {Yu}} \emph {et~al.},\ }\bibfield  {title} {\bibinfo {title} {Bepcii
  performance and beam dynamics studies on luminosity},\ }in\ \href
  {https://doi.org/10.18429/JACoW-IPAC2016-TUYA01} {\emph {\bibinfo {booktitle}
  {Proceedings of IPAC2016}}}\ (\bibinfo {address} {Busan, Korea},\ \bibinfo
  {year} {2016})\BibitemShut {NoStop}%
\bibitem [{\citenamefont {Ablikim}\ \emph {et~al.}(2020)\citenamefont {Ablikim}
  \emph {et~al.}}]{Ablikim:2019hff}%
  \BibitemOpen
  \bibfield  {author} {\bibinfo {author} {\bibfnamefont {M.}~\bibnamefont
  {Ablikim}} \emph {et~al.} (\bibinfo {collaboration} {BESIII Collaboration}),\
  }\bibfield  {title} {\bibinfo {title} {Future physics programme of besiii},\
  }\href {https://doi.org/10.1088/1674-1137/44/4/040001} {\bibfield  {journal}
  {\bibinfo  {journal} {Chin. Phys. C}\ }\textbf {\bibinfo {volume} {44}},\
  \bibinfo {pages} {1} (\bibinfo {year} {2020})}\BibitemShut {NoStop}%
\bibitem [{\citenamefont {Li}\ \emph {et~al.}(2017)\citenamefont {Li},
  \citenamefont {Yang}, \citenamefont {Zhou}, \citenamefont {Sun},
  \citenamefont {Sun}, \citenamefont {Xu}, \citenamefont {Yang}, \citenamefont
  {Yu}, \citenamefont {Zhang},\ and\ \citenamefont {Zhao}}]{etof1}%
  \BibitemOpen
  \bibfield  {author} {\bibinfo {author} {\bibfnamefont {X.}~\bibnamefont
  {Li}}, \bibinfo {author} {\bibfnamefont {Y.}~\bibnamefont {Yang}}, \bibinfo
  {author} {\bibfnamefont {L.}~\bibnamefont {Zhou}}, \bibinfo {author}
  {\bibfnamefont {S.}~\bibnamefont {Sun}}, \bibinfo {author} {\bibfnamefont
  {Z.}~\bibnamefont {Sun}}, \bibinfo {author} {\bibfnamefont {H.}~\bibnamefont
  {Xu}}, \bibinfo {author} {\bibfnamefont {J.}~\bibnamefont {Yang}}, \bibinfo
  {author} {\bibfnamefont {L.}~\bibnamefont {Yu}}, \bibinfo {author}
  {\bibfnamefont {Y.}~\bibnamefont {Zhang}},\ and\ \bibinfo {author}
  {\bibfnamefont {Y.}~\bibnamefont {Zhao}},\ }\bibfield  {title} {\bibinfo
  {title} {Measurements and characteristics of ${\rm al}_2{\rm o}_3:{\rm
  cr}^{3+}$ coating for the proton beam imaging system},\ }\href
  {https://doi.org/10.1007/s41605-017-0006-2} {\bibfield  {journal} {\bibinfo
  {journal} {Radiation Detection Technology and Methods}\ }\textbf {\bibinfo
  {volume} {1}},\ \bibinfo {pages} {13} (\bibinfo {year} {2017})}\BibitemShut
  {NoStop}%
\bibitem [{\citenamefont {Guo}\ and\ \citenamefont {{et al.}}(2017)}]{etof2}%
  \BibitemOpen
  \bibfield  {author} {\bibinfo {author} {\bibfnamefont {Y.~X.}\ \bibnamefont
  {Guo}}\ and\ \bibinfo {author} {\bibnamefont {{et al.}}},\ }\bibfield
  {title} {\bibinfo {title} {The study of time calibration for upgraded end cap
  tof of besiii},\ }\href@noop {} {\bibfield  {journal} {\bibinfo  {journal}
  {Radiation Detection Technology and Methods}\ }\textbf {\bibinfo {volume}
  {1}},\ \bibinfo {pages} {15} (\bibinfo {year} {2017})}\BibitemShut {NoStop}%
\bibitem [{\citenamefont {Cao}\ \emph {et~al.}(2020)\citenamefont {Cao},
  \citenamefont {Chen}, \citenamefont {Chen}, \citenamefont {Chen},
  \citenamefont {Chen}, \citenamefont {Ding}, \citenamefont {Fu}, \citenamefont
  {Guan}, \citenamefont {Guo}, \citenamefont {Huang}, \citenamefont {Liang},
  \citenamefont {Liu}, \citenamefont {Liu}, \citenamefont {Li}, \citenamefont
  {Li}, \citenamefont {Li}, \citenamefont {Ma}, \citenamefont {Ma},
  \citenamefont {Mao}, \citenamefont {Mo}, \citenamefont {Ping}, \citenamefont
  {Sun}, \citenamefont {Sun}, \citenamefont {Sun}, \citenamefont {Wang},
  \citenamefont {Wang}, \citenamefont {Wang}, \citenamefont {Wang},
  \citenamefont {Wei}, \citenamefont {Wu}, \citenamefont {Wu}, \citenamefont
  {Wu}, \citenamefont {Wu}, \citenamefont {Qin}, \citenamefont {Xu},
  \citenamefont {Zhang}, \citenamefont {Zhang}, \citenamefont {Zhang},
  \citenamefont {Zhang}, \citenamefont {Zhang}, \citenamefont {Zhang},
  \citenamefont {Zhang}, \citenamefont {Zheng}, \citenamefont {Zhou},
  \citenamefont {Zhou}, \citenamefont {Zhou}, \citenamefont {Zhu},
  \citenamefont {Zhu}, \citenamefont {Zhu}, \citenamefont {Zhao}, \citenamefont
  {Ma}, \citenamefont {Ma}, \citenamefont {Sun}, \citenamefont {Qiu},\ and\
  \citenamefont {Zheng}}]{etof3}%
  \BibitemOpen
  \bibfield  {author} {\bibinfo {author} {\bibfnamefont {P.}~\bibnamefont
  {Cao}}, \bibinfo {author} {\bibfnamefont {H.~F.}\ \bibnamefont {Chen}},
  \bibinfo {author} {\bibfnamefont {M.~M.}\ \bibnamefont {Chen}}, \bibinfo
  {author} {\bibfnamefont {Y.~B.}\ \bibnamefont {Chen}}, \bibinfo {author}
  {\bibfnamefont {Z.~Y.}\ \bibnamefont {Chen}}, \bibinfo {author}
  {\bibfnamefont {X.~C.}\ \bibnamefont {Ding}}, \bibinfo {author}
  {\bibfnamefont {Q.}~\bibnamefont {Fu}}, \bibinfo {author} {\bibfnamefont
  {Y.~H.}\ \bibnamefont {Guan}}, \bibinfo {author} {\bibfnamefont {Y.~X.}\
  \bibnamefont {Guo}}, \bibinfo {author} {\bibfnamefont {X.~T.}\ \bibnamefont
  {Huang}}, \bibinfo {author} {\bibfnamefont {Y.}~\bibnamefont {Liang}},
  \bibinfo {author} {\bibfnamefont {Y.~B.}\ \bibnamefont {Liu}}, \bibinfo
  {author} {\bibfnamefont {Y.}~\bibnamefont {Liu}}, \bibinfo {author}
  {\bibfnamefont {W.~G.}\ \bibnamefont {Li}}, \bibinfo {author} {\bibfnamefont
  {X.~R.}\ \bibnamefont {Li}}, \bibinfo {author} {\bibfnamefont {Z.~J.}\
  \bibnamefont {Li}}, \bibinfo {author} {\bibfnamefont {Q.~M.}\ \bibnamefont
  {Ma}}, \bibinfo {author} {\bibfnamefont {Q.~A.}\ \bibnamefont {Ma}}, \bibinfo
  {author} {\bibfnamefont {Y.~J.}\ \bibnamefont {Mao}}, \bibinfo {author}
  {\bibfnamefont {J.~S.}\ \bibnamefont {Mo}}, \bibinfo {author} {\bibfnamefont
  {J.~Q.}\ \bibnamefont {Ping}}, \bibinfo {author} {\bibfnamefont {J.~P.}\
  \bibnamefont {Sun}}, \bibinfo {author} {\bibfnamefont {X.~Y.}\ \bibnamefont
  {Sun}}, \bibinfo {author} {\bibfnamefont {Y.~J.}\ \bibnamefont {Sun}},
  \bibinfo {author} {\bibfnamefont {J.~M.}\ \bibnamefont {Wang}}, \bibinfo
  {author} {\bibfnamefont {Y.~F.}\ \bibnamefont {Wang}}, \bibinfo {author}
  {\bibfnamefont {Z.~H.}\ \bibnamefont {Wang}}, \bibinfo {author}
  {\bibfnamefont {Z.~Y.}\ \bibnamefont {Wang}}, \bibinfo {author}
  {\bibfnamefont {X.~M.}\ \bibnamefont {Wei}}, \bibinfo {author} {\bibfnamefont
  {J.~J.}\ \bibnamefont {Wu}}, \bibinfo {author} {\bibfnamefont {M.~J.}\
  \bibnamefont {Wu}}, \bibinfo {author} {\bibfnamefont {Y.}~\bibnamefont {Wu}},
  \bibinfo {author} {\bibfnamefont {Z.}~\bibnamefont {Wu}}, \bibinfo {author}
  {\bibfnamefont {Z.~H.}\ \bibnamefont {Qin}}, \bibinfo {author} {\bibfnamefont
  {L.~L.}\ \bibnamefont {Xu}}, \bibinfo {author} {\bibfnamefont {G.~F.}\
  \bibnamefont {Zhang}}, \bibinfo {author} {\bibfnamefont {H.~H.}\ \bibnamefont
  {Zhang}}, \bibinfo {author} {\bibfnamefont {J.~Y.}\ \bibnamefont {Zhang}},
  \bibinfo {author} {\bibfnamefont {L.}~\bibnamefont {Zhang}}, \bibinfo
  {author} {\bibfnamefont {S.~Q.}\ \bibnamefont {Zhang}}, \bibinfo {author}
  {\bibfnamefont {X.~M.}\ \bibnamefont {Zhang}}, \bibinfo {author}
  {\bibfnamefont {Y.}~\bibnamefont {Zhang}}, \bibinfo {author} {\bibfnamefont
  {Y.~H.}\ \bibnamefont {Zheng}}, \bibinfo {author} {\bibfnamefont {X.~K.}\
  \bibnamefont {Zhou}}, \bibinfo {author} {\bibfnamefont {Y.}~\bibnamefont
  {Zhou}}, \bibinfo {author} {\bibfnamefont {Z.~Y.}\ \bibnamefont {Zhou}},
  \bibinfo {author} {\bibfnamefont {C.~C.}\ \bibnamefont {Zhu}}, \bibinfo
  {author} {\bibfnamefont {K.~J.}\ \bibnamefont {Zhu}}, \bibinfo {author}
  {\bibfnamefont {Z.~A.}\ \bibnamefont {Zhu}}, \bibinfo {author} {\bibfnamefont
  {J.~Z.}\ \bibnamefont {Zhao}}, \bibinfo {author} {\bibfnamefont {H.~L.}\
  \bibnamefont {Ma}}, \bibinfo {author} {\bibfnamefont {X.~Y.}\ \bibnamefont
  {Ma}}, \bibinfo {author} {\bibfnamefont {Y.~Z.}\ \bibnamefont {Sun}},
  \bibinfo {author} {\bibfnamefont {J.~F.}\ \bibnamefont {Qiu}},\ and\ \bibinfo
  {author} {\bibfnamefont {X.~H.}\ \bibnamefont {Zheng}},\ }\bibfield  {title}
  {\bibinfo {title} {Design and construction of the new besiii endcap
  time-of-flight system with mrpc technology},\ }\href
  {https://doi.org/10.1016/j.nima.2019.163053} {\bibfield  {journal} {\bibinfo
  {journal} {Nuclear Instruments and Methods in Physics Research Section A:
  Accelerators, Spectrometers, Detectors and Associated Equipment}\ }\textbf
  {\bibinfo {volume} {953}},\ \bibinfo {pages} {163053} (\bibinfo {year}
  {2020})}\BibitemShut {NoStop}%
\bibitem [{\citenamefont {Yuan}\ and\ \citenamefont
  {Olsen}(2019)}]{Yuan_review}%
  \BibitemOpen
  \bibfield  {author} {\bibinfo {author} {\bibfnamefont {C.-Z.}\ \bibnamefont
  {Yuan}}\ and\ \bibinfo {author} {\bibfnamefont {S.~L.}\ \bibnamefont
  {Olsen}},\ }\bibfield  {title} {\bibinfo {title} {The besiii physics
  programme},\ }\href {https://doi.org/10.1038/s42254-019-0082-y} {\bibfield
  {journal} {\bibinfo  {journal} {Nature Reviews Physics}\ }\textbf {\bibinfo
  {volume} {1}},\ \bibinfo {pages} {480–494} (\bibinfo {year}
  {2019})}\BibitemShut {NoStop}%
\bibitem [{\citenamefont {Ablikim}\ \emph
  {et~al.}(2017{\natexlab{b}})\citenamefont {Ablikim} \emph
  {et~al.}}]{4230_pipihc}%
  \BibitemOpen
  \bibfield  {author} {\bibinfo {author} {\bibfnamefont {M.}~\bibnamefont
  {Ablikim}} \emph {et~al.} (\bibinfo {collaboration} {BESIII Collaboration}),\
  }\bibfield  {title} {\bibinfo {title} {{Evidence of Two Resonant Structures
  in $e^+ e^- \to \pi^+ \pi^- h_c$}},\ }\href
  {https://doi.org/10.1103/PhysRevLett.118.092002} {\bibfield  {journal}
  {\bibinfo  {journal} {Phys. Rev. Lett.}\ }\textbf {\bibinfo {volume} {118}},\
  \bibinfo {pages} {092002} (\bibinfo {year} {2017}{\natexlab{b}})},\ \Eprint
  {https://arxiv.org/abs/1610.07044} {arXiv:1610.07044 [hep-ex]} \BibitemShut
  {NoStop}%
\bibitem [{\citenamefont {Ablikim}\ \emph {et~al.}(2013)\citenamefont {Ablikim}
  \emph {et~al.}}]{3900_pihc}%
  \BibitemOpen
  \bibfield  {author} {\bibinfo {author} {\bibfnamefont {M.}~\bibnamefont
  {Ablikim}} \emph {et~al.} (\bibinfo {collaboration} {BESIII Collaboration}),\
  }\bibfield  {title} {\bibinfo {title} {{Observation of a Charged
  Charmoniumlike Structure ${Z}_c$(4020) and Search for the ${Z}_c$(3900) in
  $e^+e^- \to \pi^+\pi^-h_c$}},\ }\href
  {https://doi.org/10.1103/PhysRevLett.111.242001} {\bibfield  {journal}
  {\bibinfo  {journal} {Phys. Rev. Lett.}\ }\textbf {\bibinfo {volume} {111}},\
  \bibinfo {pages} {242001} (\bibinfo {year} {2013})},\ \Eprint
  {https://arxiv.org/abs/1309.1896} {arXiv:1309.1896 [hep-ex]} \BibitemShut
  {NoStop}%
\bibitem [{\citenamefont {Ablikim}\ \emph
  {et~al.}(2025{\natexlab{d}})\citenamefont {Ablikim} \emph
  {et~al.}}]{3res_pipihc}%
  \BibitemOpen
  \bibfield  {author} {\bibinfo {author} {\bibfnamefont {M.}~\bibnamefont
  {Ablikim}} \emph {et~al.} (\bibinfo {collaboration} {BESIII Collaboration}),\
  }\bibfield  {title} {\bibinfo {title} {{Observation of a Three-Resonance
  Structure in the Cross Section of $e^+e^-\to\pi^+\pi^- h_c$}},\ }\href@noop
  {} {\  (\bibinfo {year} {2025}{\natexlab{d}})},\ \Eprint
  {https://arxiv.org/abs/2504.04096} {arXiv:2504.04096 [hep-ex]} \BibitemShut
  {NoStop}%
\bibitem [{\citenamefont {Ablikim}\ \emph {et~al.}(2014)\citenamefont {Ablikim}
  \emph {et~al.}}]{pi0pi0hc}%
  \BibitemOpen
  \bibfield  {author} {\bibinfo {author} {\bibfnamefont {M.}~\bibnamefont
  {Ablikim}} \emph {et~al.} (\bibinfo {collaboration} {BESIII Collaboration}),\
  }\bibfield  {title} {\bibinfo {title} {{Observation of $e^+e^- →
  \pi^0\pi^0h_c$ and a Neutral Charmoniumlike Structure $Z_c(4020)^0$}},\
  }\href {https://doi.org/10.1103/PhysRevLett.113.212002} {\bibfield  {journal}
  {\bibinfo  {journal} {Phys. Rev. Lett.}\ }\textbf {\bibinfo {volume} {113}},\
  \bibinfo {pages} {212002} (\bibinfo {year} {2014})},\ \Eprint
  {https://arxiv.org/abs/1409.6577} {arXiv:1409.6577 [hep-ex]} \BibitemShut
  {NoStop}%
\bibitem [{\citenamefont {Ablikim}\ \emph {et~al.}(2019)\citenamefont {Ablikim}
  \emph {et~al.}}]{3pietac}%
  \BibitemOpen
  \bibfield  {author} {\bibinfo {author} {\bibfnamefont {M.}~\bibnamefont
  {Ablikim}} \emph {et~al.} (\bibinfo {collaboration} {BESIII Collaboration}),\
  }\bibfield  {title} {\bibinfo {title} {{Study of $e^+e^- \to \pi^+ \pi^-
  \pi^0 \eta_c$ and evidence for $Z_c$(3900)$^\pm$ decaying into $\rho^\pm
  \eta_c$}},\ }\href {https://doi.org/10.1103/PhysRevD.100.111102} {\bibfield
  {journal} {\bibinfo  {journal} {Phys. Rev. D}\ }\textbf {\bibinfo {volume}
  {100}},\ \bibinfo {pages} {111102} (\bibinfo {year} {2019})},\ \Eprint
  {https://arxiv.org/abs/1906.00831} {arXiv:1906.00831 [hep-ex]} \BibitemShut
  {NoStop}%
\bibitem [{\citenamefont {Rong-Gang}(2008)}]{EVTGEN}%
  \BibitemOpen
  \bibfield  {author} {\bibinfo {author} {\bibfnamefont {P.}~\bibnamefont
  {Rong-Gang}},\ }\bibfield  {title} {\bibinfo {title} {Event generators at
  {BESIII}*},\ }\href {https://doi.org/10.1088/1674-1137/32/8/001} {\bibfield
  {journal} {\bibinfo  {journal} {Chinese Physics C}\ }\textbf {\bibinfo
  {volume} {32}},\ \bibinfo {pages} {599} (\bibinfo {year} {2008})}\BibitemShut
  {NoStop}%
\bibitem [{\citenamefont {Chen}\ \emph {et~al.}(2000)\citenamefont {Chen},
  \citenamefont {Huang}, \citenamefont {Qi}, \citenamefont {Zhang},\ and\
  \citenamefont {Zhu}}]{LUNDCHARM}%
  \BibitemOpen
  \bibfield  {author} {\bibinfo {author} {\bibfnamefont {J.~C.}\ \bibnamefont
  {Chen}}, \bibinfo {author} {\bibfnamefont {G.~S.}\ \bibnamefont {Huang}},
  \bibinfo {author} {\bibfnamefont {X.~R.}\ \bibnamefont {Qi}}, \bibinfo
  {author} {\bibfnamefont {D.~H.}\ \bibnamefont {Zhang}},\ and\ \bibinfo
  {author} {\bibfnamefont {Y.~S.}\ \bibnamefont {Zhu}},\ }\bibfield  {title}
  {\bibinfo {title} {Event generator for ${J}/\ensuremath{\psi}$ and
  $\ensuremath{\psi}(2s)$ decay},\ }\href
  {https://doi.org/10.1103/PhysRevD.62.034003} {\bibfield  {journal} {\bibinfo
  {journal} {Phys. Rev. D}\ }\textbf {\bibinfo {volume} {62}},\ \bibinfo
  {pages} {034003} (\bibinfo {year} {2000})}\BibitemShut {NoStop}%
\bibitem [{\citenamefont {Kirillov}\ \emph {et~al.}(2023)\citenamefont
  {Kirillov}, \citenamefont {Mintun}, \citenamefont {Ravi}, \citenamefont
  {Mao}, \citenamefont {Rolland}, \citenamefont {Gustafson}, \citenamefont
  {Xiao}, \citenamefont {Whitehead}, \citenamefont {Berg}, \citenamefont {Lo},
  \citenamefont {Dollár},\ and\ \citenamefont {Girshick}}]{SAM}%
  \BibitemOpen
  \bibfield  {author} {\bibinfo {author} {\bibfnamefont {A.}~\bibnamefont
  {Kirillov}}, \bibinfo {author} {\bibfnamefont {E.}~\bibnamefont {Mintun}},
  \bibinfo {author} {\bibfnamefont {N.}~\bibnamefont {Ravi}}, \bibinfo {author}
  {\bibfnamefont {H.}~\bibnamefont {Mao}}, \bibinfo {author} {\bibfnamefont
  {C.}~\bibnamefont {Rolland}}, \bibinfo {author} {\bibfnamefont
  {L.}~\bibnamefont {Gustafson}}, \bibinfo {author} {\bibfnamefont
  {T.}~\bibnamefont {Xiao}}, \bibinfo {author} {\bibfnamefont {S.}~\bibnamefont
  {Whitehead}}, \bibinfo {author} {\bibfnamefont {A.~C.}\ \bibnamefont {Berg}},
  \bibinfo {author} {\bibfnamefont {W.-Y.}\ \bibnamefont {Lo}}, \bibinfo
  {author} {\bibfnamefont {P.}~\bibnamefont {Dollár}},\ and\ \bibinfo {author}
  {\bibfnamefont {R.}~\bibnamefont {Girshick}},\ }\href
  {https://arxiv.org/abs/2304.02643} {\bibinfo {title} {Segment anything}}
  (\bibinfo {year} {2023}),\ \Eprint {https://arxiv.org/abs/2304.02643}
  {arXiv:2304.02643 [cs.CV]} \BibitemShut {NoStop}%
\bibitem [{\citenamefont {Jones}\ \emph {et~al.}(1998)\citenamefont {Jones},
  \citenamefont {Schonlau},\ and\ \citenamefont {Welch}}]{bayes}%
  \BibitemOpen
  \bibfield  {author} {\bibinfo {author} {\bibfnamefont {D.~R.}\ \bibnamefont
  {Jones}}, \bibinfo {author} {\bibfnamefont {M.}~\bibnamefont {Schonlau}},\
  and\ \bibinfo {author} {\bibfnamefont {W.~J.}\ \bibnamefont {Welch}},\
  }\bibfield  {title} {\bibinfo {title} {Efficient global optimization of
  expensive black-box functions},\ }\href
  {https://doi.org/10.1023/A:1008306431147} {\bibfield  {journal} {\bibinfo
  {journal} {Journal of Global Optimization}\ }\textbf {\bibinfo {volume}
  {13}},\ \bibinfo {pages} {455} (\bibinfo {year} {1998})}\BibitemShut
  {NoStop}%
\bibitem [{\citenamefont {Paszke}\ \emph {et~al.}(2019)\citenamefont {Paszke},
  \citenamefont {Gross}, \citenamefont {Massa}, \citenamefont {Lerer},
  \citenamefont {Bradbury}, \citenamefont {Chanan}, \citenamefont {Killeen},
  \citenamefont {Lin}, \citenamefont {Gimelshein}, \citenamefont {Antiga},
  \citenamefont {Desmaison}, \citenamefont {Köpf}, \citenamefont {Yang},
  \citenamefont {DeVito}, \citenamefont {Raison}, \citenamefont {Tejani},
  \citenamefont {Chilamkurthy}, \citenamefont {Steiner}, \citenamefont {Fang},
  \citenamefont {Bai},\ and\ \citenamefont {Chintala}}]{pytorch}%
  \BibitemOpen
  \bibfield  {author} {\bibinfo {author} {\bibfnamefont {A.}~\bibnamefont
  {Paszke}}, \bibinfo {author} {\bibfnamefont {S.}~\bibnamefont {Gross}},
  \bibinfo {author} {\bibfnamefont {F.}~\bibnamefont {Massa}}, \bibinfo
  {author} {\bibfnamefont {A.}~\bibnamefont {Lerer}}, \bibinfo {author}
  {\bibfnamefont {J.}~\bibnamefont {Bradbury}}, \bibinfo {author}
  {\bibfnamefont {G.}~\bibnamefont {Chanan}}, \bibinfo {author} {\bibfnamefont
  {T.}~\bibnamefont {Killeen}}, \bibinfo {author} {\bibfnamefont
  {Z.}~\bibnamefont {Lin}}, \bibinfo {author} {\bibfnamefont {N.}~\bibnamefont
  {Gimelshein}}, \bibinfo {author} {\bibfnamefont {L.}~\bibnamefont {Antiga}},
  \bibinfo {author} {\bibfnamefont {A.}~\bibnamefont {Desmaison}}, \bibinfo
  {author} {\bibfnamefont {A.}~\bibnamefont {Köpf}}, \bibinfo {author}
  {\bibfnamefont {E.}~\bibnamefont {Yang}}, \bibinfo {author} {\bibfnamefont
  {Z.}~\bibnamefont {DeVito}}, \bibinfo {author} {\bibfnamefont
  {M.}~\bibnamefont {Raison}}, \bibinfo {author} {\bibfnamefont
  {A.}~\bibnamefont {Tejani}}, \bibinfo {author} {\bibfnamefont
  {S.}~\bibnamefont {Chilamkurthy}}, \bibinfo {author} {\bibfnamefont
  {B.}~\bibnamefont {Steiner}}, \bibinfo {author} {\bibfnamefont
  {L.}~\bibnamefont {Fang}}, \bibinfo {author} {\bibfnamefont {J.}~\bibnamefont
  {Bai}},\ and\ \bibinfo {author} {\bibfnamefont {S.}~\bibnamefont
  {Chintala}},\ }\href {https://arxiv.org/abs/1912.01703} {\bibinfo {title}
  {Pytorch: An imperative style, high-performance deep learning library}}
  (\bibinfo {year} {2019}),\ \Eprint {https://arxiv.org/abs/1912.01703}
  {arXiv:1912.01703 [cs.LG]} \BibitemShut {NoStop}%
\bibitem [{\citenamefont {Kendall}\ \emph {et~al.}(2018)\citenamefont
  {Kendall}, \citenamefont {Gal},\ and\ \citenamefont {Cipolla}}]{UW}%
  \BibitemOpen
  \bibfield  {author} {\bibinfo {author} {\bibfnamefont {A.}~\bibnamefont
  {Kendall}}, \bibinfo {author} {\bibfnamefont {Y.}~\bibnamefont {Gal}},\ and\
  \bibinfo {author} {\bibfnamefont {R.}~\bibnamefont {Cipolla}},\ }\href
  {https://arxiv.org/abs/1705.07115} {\bibinfo {title} {Multi-task learning
  using uncertainty to weigh losses for scene geometry and semantics}}
  (\bibinfo {year} {2018}),\ \Eprint {https://arxiv.org/abs/1705.07115}
  {arXiv:1705.07115 [cs.CV]} \BibitemShut {NoStop}%
\bibitem [{\citenamefont {Qiu}\ \emph {et~al.}(2025)\citenamefont {Qiu},
  \citenamefont {Wang}, \citenamefont {Zheng}, \citenamefont {Huang},
  \citenamefont {Wen}, \citenamefont {Yang}, \citenamefont {Men}, \citenamefont
  {Yu}, \citenamefont {Huang}, \citenamefont {Huang}, \citenamefont {Liu},
  \citenamefont {Zhou},\ and\ \citenamefont {Lin}}]{gated_attn}%
  \BibitemOpen
  \bibfield  {author} {\bibinfo {author} {\bibfnamefont {Z.}~\bibnamefont
  {Qiu}}, \bibinfo {author} {\bibfnamefont {Z.}~\bibnamefont {Wang}}, \bibinfo
  {author} {\bibfnamefont {B.}~\bibnamefont {Zheng}}, \bibinfo {author}
  {\bibfnamefont {Z.}~\bibnamefont {Huang}}, \bibinfo {author} {\bibfnamefont
  {K.}~\bibnamefont {Wen}}, \bibinfo {author} {\bibfnamefont {S.}~\bibnamefont
  {Yang}}, \bibinfo {author} {\bibfnamefont {R.}~\bibnamefont {Men}}, \bibinfo
  {author} {\bibfnamefont {L.}~\bibnamefont {Yu}}, \bibinfo {author}
  {\bibfnamefont {F.}~\bibnamefont {Huang}}, \bibinfo {author} {\bibfnamefont
  {S.}~\bibnamefont {Huang}}, \bibinfo {author} {\bibfnamefont
  {D.}~\bibnamefont {Liu}}, \bibinfo {author} {\bibfnamefont {J.}~\bibnamefont
  {Zhou}},\ and\ \bibinfo {author} {\bibfnamefont {J.}~\bibnamefont {Lin}},\
  }\href {https://arxiv.org/abs/2505.06708} {\bibinfo {title} {Gated attention
  for large language models: Non-linearity, sparsity, and attention-sink-free}}
  (\bibinfo {year} {2025}),\ \Eprint {https://arxiv.org/abs/2505.06708}
  {arXiv:2505.06708 [cs.CL]} \BibitemShut {NoStop}%
\bibitem [{\citenamefont {Peebles}\ and\ \citenamefont {Xie}(2023)}]{CLN}%
  \BibitemOpen
  \bibfield  {author} {\bibinfo {author} {\bibfnamefont {W.}~\bibnamefont
  {Peebles}}\ and\ \bibinfo {author} {\bibfnamefont {S.}~\bibnamefont {Xie}},\
  }\href {https://arxiv.org/abs/2212.09748} {\bibinfo {title} {Scalable
  diffusion models with transformers}} (\bibinfo {year} {2023}),\ \Eprint
  {https://arxiv.org/abs/2212.09748} {arXiv:2212.09748 [cs.CV]} \BibitemShut
  {NoStop}%
\bibitem [{\citenamefont {Kingma}\ and\ \citenamefont {Ba}(2017)}]{Adam}%
  \BibitemOpen
  \bibfield  {author} {\bibinfo {author} {\bibfnamefont {D.~P.}\ \bibnamefont
  {Kingma}}\ and\ \bibinfo {author} {\bibfnamefont {J.}~\bibnamefont {Ba}},\
  }\href {https://arxiv.org/abs/1412.6980} {\bibinfo {title} {Adam: A method
  for stochastic optimization}} (\bibinfo {year} {2017}),\ \Eprint
  {https://arxiv.org/abs/1412.6980} {arXiv:1412.6980 [cs.LG]} \BibitemShut
  {NoStop}%
\bibitem [{\citenamefont {Hsu}\ \emph {et~al.}(2026)\citenamefont {Hsu},
  \citenamefont {Zhou}, \citenamefont {Liu}, \citenamefont {Xu}, \citenamefont
  {Li}, \citenamefont {Hou}, \citenamefont {Nachman}, \citenamefont {Hsu},
  \citenamefont {Mikuni}, \citenamefont {Chou},\ and\ \citenamefont
  {Zhang}}]{evenet}%
  \BibitemOpen
  \bibfield  {author} {\bibinfo {author} {\bibfnamefont {T.-H.}\ \bibnamefont
  {Hsu}}, \bibinfo {author} {\bibfnamefont {B.-H.}\ \bibnamefont {Zhou}},
  \bibinfo {author} {\bibfnamefont {Q.}~\bibnamefont {Liu}}, \bibinfo {author}
  {\bibfnamefont {Y.}~\bibnamefont {Xu}}, \bibinfo {author} {\bibfnamefont
  {S.}~\bibnamefont {Li}}, \bibinfo {author} {\bibfnamefont {G.~W.-S.}\
  \bibnamefont {Hou}}, \bibinfo {author} {\bibfnamefont {B.}~\bibnamefont
  {Nachman}}, \bibinfo {author} {\bibfnamefont {S.-C.}\ \bibnamefont {Hsu}},
  \bibinfo {author} {\bibfnamefont {V.}~\bibnamefont {Mikuni}}, \bibinfo
  {author} {\bibfnamefont {Y.-T.}\ \bibnamefont {Chou}},\ and\ \bibinfo
  {author} {\bibfnamefont {Y.}~\bibnamefont {Zhang}},\ }\href
  {https://arxiv.org/abs/2601.17126} {\bibinfo {title} {Evenet: A foundation
  model for particle collision data analysis}} (\bibinfo {year} {2026}),\
  \Eprint {https://arxiv.org/abs/2601.17126} {arXiv:2601.17126 [hep-ex]}
  \BibitemShut {NoStop}%
\bibitem [{\citenamefont {Bhimji}\ \emph {et~al.}(2026)\citenamefont {Bhimji},
  \citenamefont {Harris}, \citenamefont {Mikuni},\ and\ \citenamefont
  {Nachman}}]{omniLearned}%
  \BibitemOpen
  \bibfield  {author} {\bibinfo {author} {\bibfnamefont {W.}~\bibnamefont
  {Bhimji}}, \bibinfo {author} {\bibfnamefont {C.}~\bibnamefont {Harris}},
  \bibinfo {author} {\bibfnamefont {V.}~\bibnamefont {Mikuni}},\ and\ \bibinfo
  {author} {\bibfnamefont {B.}~\bibnamefont {Nachman}},\ }\bibfield  {title}
  {\bibinfo {title} {Foundation model framework for all tasks involving jet
  physics},\ }\bibfield  {journal} {\bibinfo  {journal} {Physical Review D}\
  }\textbf {\bibinfo {volume} {113}},\ \href
  {https://doi.org/10.1103/knmd-f5jm} {10.1103/knmd-f5jm} (\bibinfo {year}
  {2026})\BibitemShut {NoStop}%
\bibitem [{\citenamefont {Elsharkawy}\ \emph {et~al.}(2026)\citenamefont
  {Elsharkawy}, \citenamefont {Mikuni}, \citenamefont {Bhimji},\ and\
  \citenamefont {Nachman}}]{omnimol}%
  \BibitemOpen
  \bibfield  {author} {\bibinfo {author} {\bibfnamefont {I.}~\bibnamefont
  {Elsharkawy}}, \bibinfo {author} {\bibfnamefont {V.}~\bibnamefont {Mikuni}},
  \bibinfo {author} {\bibfnamefont {W.}~\bibnamefont {Bhimji}},\ and\ \bibinfo
  {author} {\bibfnamefont {B.}~\bibnamefont {Nachman}},\ }\href
  {https://arxiv.org/abs/2601.10791} {\bibinfo {title} {Omnimol: Transferring
  particle physics knowledge to molecular dynamics with point-edge
  transformers}} (\bibinfo {year} {2026}),\ \Eprint
  {https://arxiv.org/abs/2601.10791} {arXiv:2601.10791 [physics.chem-ph]}
  \BibitemShut {NoStop}%
\bibitem [{\citenamefont {Barrué}\ \emph {et~al.}(2026)\citenamefont
  {Barrué}, \citenamefont {Benato}, \citenamefont {Güven}, \citenamefont
  {Hammou}, \citenamefont {ter Hoeve}, \citenamefont {Krause}, \citenamefont
  {Li}, \citenamefont {Mantani}, \citenamefont {Rojo}, \citenamefont {Cruz},
  \citenamefont {Schöfbeck}, \citenamefont {Ubiali},\ and\ \citenamefont
  {Wang}}]{Inference1}%
  \BibitemOpen
  \bibfield  {author} {\bibinfo {author} {\bibfnamefont {R.}~\bibnamefont
  {Barrué}}, \bibinfo {author} {\bibfnamefont {L.}~\bibnamefont {Benato}},
  \bibinfo {author} {\bibfnamefont {A.~K.}\ \bibnamefont {Güven}}, \bibinfo
  {author} {\bibfnamefont {E.}~\bibnamefont {Hammou}}, \bibinfo {author}
  {\bibfnamefont {J.}~\bibnamefont {ter Hoeve}}, \bibinfo {author}
  {\bibfnamefont {C.}~\bibnamefont {Krause}}, \bibinfo {author} {\bibfnamefont
  {A.}~\bibnamefont {Li}}, \bibinfo {author} {\bibfnamefont {L.}~\bibnamefont
  {Mantani}}, \bibinfo {author} {\bibfnamefont {J.}~\bibnamefont {Rojo}},
  \bibinfo {author} {\bibfnamefont {S.~S.}\ \bibnamefont {Cruz}}, \bibinfo
  {author} {\bibfnamefont {R.}~\bibnamefont {Schöfbeck}}, \bibinfo {author}
  {\bibfnamefont {M.}~\bibnamefont {Ubiali}},\ and\ \bibinfo {author}
  {\bibfnamefont {D.}~\bibnamefont {Wang}},\ }\href
  {https://arxiv.org/abs/2604.13157} {\bibinfo {title} {Proton structure from
  neural simulation-based inference at the lhc}} (\bibinfo {year} {2026}),\
  \Eprint {https://arxiv.org/abs/2604.13157} {arXiv:2604.13157 [hep-ph]}
  \BibitemShut {NoStop}%
\bibitem [{\citenamefont {Woodward}\ \emph {et~al.}(2026)\citenamefont
  {Woodward}, \citenamefont {Villarreal}, \citenamefont {Hardin}, \citenamefont
  {Schneider},\ and\ \citenamefont {Conrad}}]{Inference2}%
  \BibitemOpen
  \bibfield  {author} {\bibinfo {author} {\bibfnamefont {J.~P.}\ \bibnamefont
  {Woodward}}, \bibinfo {author} {\bibfnamefont {J.}~\bibnamefont
  {Villarreal}}, \bibinfo {author} {\bibfnamefont {J.~M.}\ \bibnamefont
  {Hardin}}, \bibinfo {author} {\bibfnamefont {A.}~\bibnamefont {Schneider}},\
  and\ \bibinfo {author} {\bibfnamefont {J.~M.}\ \bibnamefont {Conrad}},\
  }\href {https://arxiv.org/abs/2603.15322} {\bibinfo {title} {A
  simulation-based inference evaluation of tension between microboone and
  miniboone results in a 3+1 sterile neutrino global fit}} (\bibinfo {year}
  {2026}),\ \Eprint {https://arxiv.org/abs/2603.15322} {arXiv:2603.15322
  [hep-ex]} \BibitemShut {NoStop}%
\bibitem [{\citenamefont {Tame-Narvaez}\ \emph {et~al.}(2026)\citenamefont
  {Tame-Narvaez}, \citenamefont {Gardiner}, \citenamefont {Ćiprijanović},\
  and\ \citenamefont {Cerati}}]{Inference3}%
  \BibitemOpen
  \bibfield  {author} {\bibinfo {author} {\bibfnamefont {K.}~\bibnamefont
  {Tame-Narvaez}}, \bibinfo {author} {\bibfnamefont {S.}~\bibnamefont
  {Gardiner}}, \bibinfo {author} {\bibfnamefont {A.}~\bibnamefont
  {Ćiprijanović}},\ and\ \bibinfo {author} {\bibfnamefont {G.}~\bibnamefont
  {Cerati}},\ }\href {https://arxiv.org/abs/2603.09778} {\bibinfo {title}
  {First estimation of model parameters for neutrino-induced nucleon knockout
  using simulation-based inference}} (\bibinfo {year} {2026}),\ \Eprint
  {https://arxiv.org/abs/2603.09778} {arXiv:2603.09778 [hep-ph]} \BibitemShut
  {NoStop}%
\end{thebibliography}%

\end{document}